\documentclass[12pt]{article}

\usepackage{fullpage}
\usepackage{subfigure}
\usepackage{latexsym}
\usepackage{graphicx}
\usepackage[intlimits]{amsmath}
\usepackage{amsthm}
\usepackage{natbib}
\usepackage{amsfonts}
\usepackage{ulem}
\usepackage{booktabs}
\usepackage{color}
\usepackage{placeins}
\usepackage{hyperref}
\usepackage{url}
\usepackage{setspace}

\newcommand{\E}{\mathop{\rm{E}\mathstrut}\nolimits}

\begin{document}

\begin{center}
   {\LARGE Quantifying and mitigating the effect of preferential sampling on phylodynamic inference}\\\ \\
  { Michael D. Karcher\textsuperscript{1}, Julia A. Palacios\textsuperscript{2,3,4},
  Trevor Bedford\textsuperscript{5}, Marc A. Suchard\textsuperscript{6,7,8},\\
  Vladimir N. Minin\textsuperscript{1,9}}\\
  {\footnotesize \textsuperscript{1}Department of Statistics, University of Washington, Seattle\\
  \textsuperscript{2}Department of Organismic and Evolutionary Biology, Harvard University \\
  \textsuperscript{3}Department of Ecology and Evolutionary Biology, Brown University\\
  \textsuperscript{4}Center for Computational Molecular Biology, Brown University\\
  \textsuperscript{5}Vaccine and Infectious Disease Division, Fred Hutchinson Cancer Research Center\\
   \textsuperscript{6}Department of Human Genetics, David Geffen School of Medicine at UCLA,\\
   University of California, Los Angeles\\
   \textsuperscript{7}Department of Biomathematics, David Geffen School of Medicine at UCLA,\\
   University of California,  Los Angeles\\
      \textsuperscript{8}Department of Biostatistics, UCLA Fielding School of Public Health,\\
      University of California, Los Angeles\\
     \textsuperscript{9}Department of Biology, University of Washington, Seattle}
\end{center}

\begin{abstract}
Phylodynamics seeks to estimate effective population size fluctuations from molecular sequences of individuals sampled from a population of interest.
One way to accomplish this task formulates an observed sequence data likelihood exploiting a coalescent model for the sampled individuals' genealogy and then integrating over all possible genealogies via Monte Carlo or, less efficiently, by conditioning on one genealogy estimated from the sequence data.
However, when analyzing sequences sampled serially through time, current methods implicitly assume either that sampling times are fixed deterministically by the data collection protocol or that their distribution does not depend on the size of the population.
Through simulation, we first show that, when sampling times do probabilistically depend on effective population size, estimation methods may be systematically biased.
To correct for this deficiency, we propose a new model that explicitly accounts for preferential sampling
by modeling the sampling times as an inhomogeneous Poisson process dependent on effective population size.
We demonstrate that in the presence of preferential sampling our new model not only reduces bias,
but also improves estimation precision.
Finally, we compare the performance of the currently used phylodynamic methods with our proposed model through clinically-relevant, seasonal human influenza examples.
\end{abstract}


\section*{Introduction}

Phylodynamics --- a set of techniques for estimating population dynamics from genetic data --- has proven useful in ecology and epidemiology \citep{grenfell2004unifying, holmes2009discovering}.
Phylodynamics is especially useful in cases where ascertaining population sizes via traditional sampling methods is infeasible; e.g., in infectious disease epidemiology it is impossible to obtain the total number of infected individuals in a large population.
Estimating population dynamics from a limited sample of genetic data is possible because changes in population size leave evidence in the molecular sequences of the population.
Recently, techniques employing a nonparametric approach to inferring population trajectories have improved upon earlier models in terms of flexibility, accuracy, and precision by, e.g., employing Gaussian Markov random fields \citep{skyride, skygrid} and Gaussian processes \citep{palacios2013gaussian}.
However, none of these state-of-the-art methods currently account for randomness in sampling time data, potentially introducing bias in studies where sampling times have a relationship to population dynamics.
Through a simulation study we characterize this bias in a demographic scenario with seasonally varying population size.
We also extend the state-of-the-art by incorporating a sampling time model into phylodynamic inference, mitigating the bias and improving precision.

Phylodynamic methods use Kingman's coalescent model that, given a particular effective population size trajectory, defines the density of a  genealogy relating the sampled individuals \citep{coalescent}.
Effective population size measures genetic diversity present in the population and relates to census population size if certain assumptions are met \citep{wakeley2009extensions}.
Many early coalescent-based phylodynamic methods required strict parametric assumptions about the effective population size trajectory, such as constant through time \citep{griffiths1994sampling} or exponential growth \citep{drummond2002estimating, kuhner1998maximum}.
A major alternative arose with the advent of nonparametric methods, one of the earliest and most influential being the piecewise constant classical skyline model \citep{pybus2000integrated}.
This approach greatly increases the number of estimated parameters, leading to noisy effective population size trajectories.
A number of algorithms seeking compromise between the relative stability of parametric approaches and the flexibility of nonparametric approaches have been implemented \citep{drummond2005bayesian, skyride, skygrid}.
For a detailed comparison, see a review by \citet{ho2011skyline}.

Many successful applications of phylodynamics methodology come from infectious disease epidemiology, where the effective population size is interpreted, albeit with caution, as the effective number of infections \citep{frost2010viral}.
In these epidemiological applications, disease agent DNA or RNA sequences are collected at multiple times.
When analyzing such heterochronous data, researchers implicitly assume that sampling times are either fixed or follow a distribution that is functionally independent of the effective population size trajectory.
However, it is conceivable that the infectious disease agent DNA samples are collected more frequently when the number of infections is high and less frequently during time periods with few infections.
Therefore, the implicit assumption of no relationship between sampling times and population dynamics, made by all state-of-the-art phylodynamic methods, is troublesome, since unrecognized preferential sampling leads to systematic estimation bias, as explored by \citet{diggle2010geostatistical} in the context of spatial statistics.
Furthermore, preferential sampling could be present in the sequence databases, but it could also be introduced accidentally or intentionally by filtering during database queries or data mining.

To test the effect of preferential sampling on phylodynamic inference we first perform a simulation study.
We simulate sampling times according to multiple distributions, contrasting distributions functionally dependent to effective population size with a functionally independent distribution.
We then simulate genealogies based on the sampling times and perform state-of-the-art phylodynamic analyses, and we find that ignoring preferential sampling can bias effective population size estimation and that the size of the bias depends on the local properties of the effective population size trajectory.

In order to account for preferential sampling, we formulate a new phylodynamic model in which sampling times are generated from an inhomogeneous Poisson process with intensity functionally dependent on effective population size.
We incorporate our Poisson preferential sampling model into a Gaussian process-based Bayesian phylodynamic method \citep{skyride, skygrid, palacios2013gaussian}.
Applying our new sampling-aware method to our simulations shows that modeling preferential sampling eliminates the aforementioned bias and can increase precision of the phylodynamic inference.
In all of our developments, we assume that the genealogy of the sample is known without error.
This assumption allows us to use an integrated nested Laplace approximation (INLA) to make our Bayesian inference computationally efficient \citep{rue2009approximate, palacios2012INLA}, which is important for executing our simulation studies.

Finally, we examine the performance of our algorithm on two real-world examples.
\citet{rambaut2008genomic} explore the seasonal variation of genetic diversity in the genes that code for several of the most important proteins in the two most common influenza subtypes, H3N2 and H1N1.
For the sake of brevity we only analyze the hemagglutinin gene in H3N2.
We find evidence of preferential sampling in the dataset, and our sampling-aware method produces a large improvement in precision over the conditional (sampling un-aware) method.
\citet{zinder2014seasonality} specifically explore the patterns of seasonal migration of genetic diversity of H3N2 influenza across the regions of the world.
We examine the regions separately and find differing strengths of preferential sampling,
but in all regions our method performs better than the conditional model.
In some regions, we see stronger relationships between sampling frequency and population size,
most often in regions with the most seasonal variation in incidence.

\section*{Methods}

\subsection*{State-of-the-art phylodynamics}

\begin{figure}[htbp]
	\centering
		\includegraphics[width=1.0\textwidth]{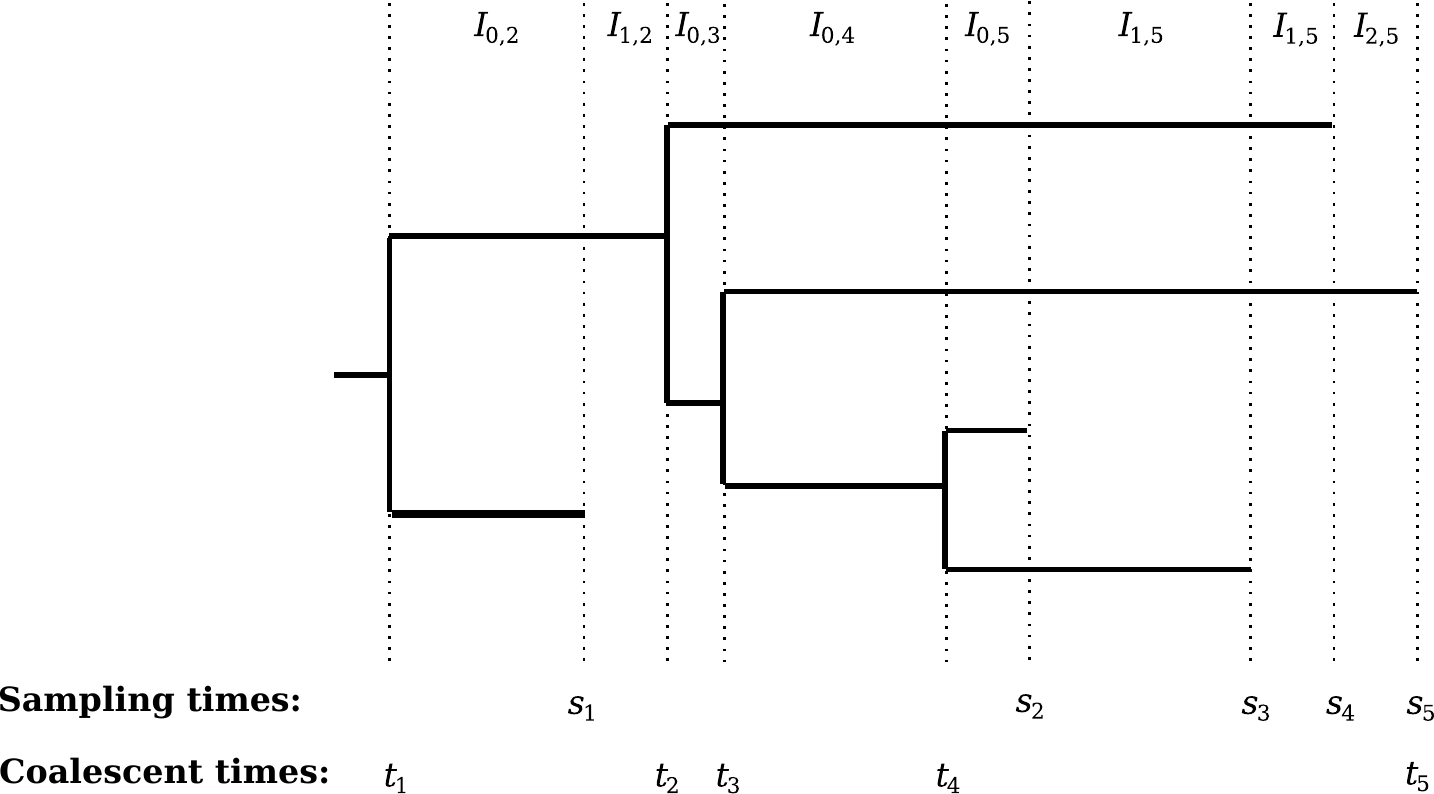}
	\caption{Illustration of an example heterochronous genealogy with $n = 5$ lineages.
		Sampling times $s_1,\ldots,s_5$ and coalescent times $t_1,\ldots,t_5$ are marked below the genealogy.}
	\label{fig:Genealogy}
\end{figure}

Consider a sample of individuals from a well-mixed population.
Some individuals will share a common ancestor more recently than others.
One pair of individuals in particular will have the pairwise most recent common ancestor.
Moving backwards in time, we can consider those two individuals to have \textit{coalesced}, treating the two individuals as one.
We can then repeat this process of finding the pairwise most recent common ancestor and coalescing individuals until we reach the most recent common ancestor of the entire sample.
If we keep track of the ancestral lineages and coalescences of the individuals, we see the data take the shape of a bifurcating tree, and we refer to this ancestry tree as a \textit{genealogy} (illustrated in Figure \ref{fig:Genealogy}).
We refer to the branching points of the genealogy tree as \textit{coalescent events}.
If the samples are all taken simultaneously, we refer to the genealogy as \textit{isochronous}.
\citet{coalescent}'s original coalescent provided a density for isochronous genealogies with a fixed effective population size.
Later extensions to the coalescent allowed for parametric and nonparametric specifications of effective population size trajectories along with \textit{heterochronous} sampling times.
Heterochronous sampling times (also called sampling events) can occur at any time up to the present.

We consider first the case of a fixed, heterochronous genealogy \citep{joseph_coalescent_1999}. The coalescent likelihood has sufficient statistics
${\mathbf{g} = \left\{ t_i \right\}_{i=1}^n}, \allowbreak {0 = t_n < t_{n-1} < \ldots < t_1}$, representing the coalescent times, and
${\mathbf{s} = \left\{ s_i, n_i \right\}_{i=1}^m}, \allowbreak {0 = s_m < s_{m-1} < \ldots < s_1}, \allowbreak {\sum_{j=1}^m {n_j} = n}$,
representing the sampling times along with the corresponding number of lineages sampled
(see Figure \ref{fig:Genealogy}).
We define the number of \textit{active lineages} at time $t$ as the number of lineages sampled between $t$ and the present, minus the number of coalescent events between $t$ and the present.
In Figure \ref{fig:Genealogy}, this appears as the number of horizontal lines that a vertical line at time $t$ will cross.
We define a partition of $(0,t_1)$ with intervals $I_{i,k}$ for $k = 1, \ldots, n$.
We let $I_{0,k}$ represent the intervals ending with a coalescent event and let $I_{i,k}$ for $i = 1, \ldots, m_k$ represent the $m_k$ intervals ending in a sampling event between the $(k-1)$th and $k$th coalescent events (see Intervals in Figure \ref{fig:Genealogy}).
We let $C_{i,k} = \binom{n_{i,k}}{2}$, where $n_{i,k}$ is the number of active lineages in the interval $I_{i,k}$.
Suppose  $\mathbf{s}$ is fixed, then the coalescent likelihood is
\[
 \Pr[\mathbf{g} | N_e(t), \mathbf{s}] \propto \prod_{k=2}^{n} {\frac{C_{0,k}}{N_e(t_{k-1})}} \exp\left[ -\sum_{i=0}^{m_k} {\int_{I_{i,k}} {\frac{C_{i,k}}{N_e(t)} dt}} \right] .
\]
\par
In Bayesian phylodynamic inference, our aim is to explore the posterior distribution of $N_e(t)$,
so we employ a Gaussian process prior $\Pr[N_e(t) \mid \tau]$ as in \citet{palacios2012INLA},
where $N_e(t) = \exp[f(t)]$, with $f(t) \sim \mathcal{BM}(\tau)$ following a Brownian motion with precision parameter $\tau$.
We assign a Gamma(0.01, 0.01) hyperprior to $\tau$.
This results in the posterior $\Pr[N_e(t), \tau \mid \mathbf{g}] \propto \Pr[\mathbf{g} \mid N_e(t)] \Pr[N_e(t) \mid \tau] \Pr(\tau)$.

The continuous case as written above involves an infinite-dimensional object---the function $N_e(t)$---which makes the problem as stated intractable.
However, we can approximate the continuous function with a piecewise constant function.
We construct a fine, regular grid $\mathbf{x} = \{ x_j \}_{j=1}^B$ with grid width $w$ over the interval that supports the genealogy and let $\gamma_j = \log[N_e(x_j)]$.
We construct a piecewise constant approximation $N^{\boldsymbol{\gamma}}(t) = \sum_{i=1}^B { \exp(\gamma_i) 1_{ t \in [ x_i - w/2, x_i + w/2 ) }}$.
The discretized coalescent likelihood becomes
\begin{equation}
 \Pr(\mathbf{g} \mid \boldsymbol{\gamma}) \propto \prod_{k=2}^{n} {\frac{C_{0,k}}{N^{\boldsymbol{\gamma}}(t_{k-1})}} \exp\left[ -\sum_{i=0}^{m_k} {\int_{I_{i,k}} {\frac{C_{i,k}}{N^{\boldsymbol{\gamma}}(t)} dt}} \right],
\label{coal-like}
\end{equation}
where $\boldsymbol{\gamma} = (\gamma_1,\ldots,\gamma_B)$ and the integrals are simple to compute over the step function $N^{\boldsymbol{\gamma}}(t)$.
We discretize the Brownian process prior with an intrinsic random walk prior,
\[
 \Pr(\boldsymbol{\gamma} \mid \tau) \propto \tau^{(n-1)/2} \exp{\left[ -\frac{\tau}{2} \sum_{k=1}^{B-1}{(\gamma_{k+1} - \gamma_k)^2} \right]}.
\]
Finally, the discretized posterior becomes $\Pr(\boldsymbol{\gamma}, \tau \mid \mathbf{g})\propto \Pr(\mathbf{g} \mid \boldsymbol{\gamma}) \Pr(\boldsymbol{\gamma} | \tau) \Pr(\tau)$.

With the posterior known (up to a proportionality constant),
we can proceed with numerical integration techniques such as Markov chain Monte Carlo (MCMC)
or INLA --- a deterministic algorithm for approximating posterior distributions.
We select INLA and name the implementation Bayesian nonparametric phylodynamic reconstruction (BNPR).

\subsection*{Phylodynamics with preferential sampling}

In the previous section we made the assumption that we could safely ignore any potential dependence of sampling times $\mathbf{s}$ on effective population size $N^{\boldsymbol{\gamma}}(t)$ in our calculations.
In this section, we relax this assumption.
We model sampling times according to an inhomogeneous Poisson process in a fixed sampling window $[0,s_0]$,
with intensity $\lambda(t) = \beta_0 [N^{\boldsymbol{\gamma}}(t)]^{\beta_1}$,
i.e. proportional to a power of the effective population size,
where $\beta_0$ and $\beta_1$ are unknown parameters.
The sampling log-likelihood is
\[
\log[\Pr(\mathbf{s}\mid \boldsymbol{\gamma}, \beta_0, \beta_1)] = C + n \log{\beta_0} + \sum_{i=1}^{n}{\beta_1 \log[N^{\boldsymbol{\gamma}}(s_i)] - \int_{s_m}^{s_0}{\beta_0 [N^{\boldsymbol{\gamma}}(r)]^{\beta_1} dr}}.
\]
To illustrate our parameterization, sampling with $\beta_1 = 1$ would result in collecting genetic sequences with intensity directly proportional to effective population size, while higher $\beta_1$ values result in more clustered samples.
Conversely, $\beta_1 = 0$ produces a uniform distribution of sampling times, with a Poisson distribution on the number of individuals sampled.

In many datasets, the sampling time data will have low enough resolution
(for instance, only recording the date but not time of sampling) that some sampling times will be appear coincident.
Our sampling model is compatible with simultaneous sampling times because the model naturally buckets the samples along our earlier discretization.
The likelihood is proportional to a product of Poisson mass functions centered at the grid points $\mathbf{x}$.

The genealogy depends on the sampling times, so we condition on $\mathbf{s}$ in the likelihood for $\mathbf{g}$.
We are treating $\mathbf{s}$ as random, so we insert the likelihood term for it as well as independent Normal priors for parameters $\beta_0$ and $\beta_1$---specifically $\beta_i \sim N(\text{mean} = 0, \text{variance} = 1000)$ for $i=0,1$.
We retain the same hyperprior for the precision parameter $\tau$ as above.
This results in the posterior that accounts for preferential sampling,
\[
\Pr(\boldsymbol{\gamma}, \tau, \boldsymbol{\beta} \mid \mathbf{g}, \mathbf{s}) \propto \Pr(\mathbf{g} \mid \mathbf{s}, \boldsymbol{\gamma}) \Pr(\mathbf{s} \mid \boldsymbol{\gamma}, \boldsymbol{\beta}) \Pr(\boldsymbol{\gamma} \mid \tau) \Pr(\tau) \Pr(\boldsymbol{\beta}),
\]
where $\Pr(\mathbf{g} \mid \mathbf{s}, \boldsymbol{\gamma})$ is defined by Equation \eqref{coal-like}, but now we add conditioning on $\mathbf{s}$ explicitly.
In the case where the sampling information $\mathbf{s}$ is functionally independent of the vector of log effective population sizes $\boldsymbol{\gamma}$, 
the posterior for $\mathbf{g}$ simplifies to the form it had in the previous section,
because the likelihood for $\mathbf{s}$ becomes a constant in $\boldsymbol{\gamma}$.
We incorporate our sampling model into an INLA framework similar to BNPR and name the implementation Bayesian nonparametric phylodynamic reconstruction with preferential sampling  (BNPR-PS).

\subsection*{INLA framework}

Here we present a brief outline of the INLA methodology \citep{rue2009approximate} in the context of our BNPR and BNPR-PS implementations.
We first examine BNPR as the simpler model.
In the end, we intend to estimate the marginal posteriors $\Pr(\tau \mid \mathbf{g})$ and $\Pr(\gamma_i \mid \mathbf{g}), i=1,\ldots,B$, most often focusing on the posterior medians and the end points of the 95\% Bayesian credible intervals.
We approximate the marginal of $\tau$ with
\[
\widehat{\Pr}(\tau \mid \mathbf{g}) \propto \left. \frac{\Pr(\boldsymbol{\gamma}, \tau, \mathbf{g})}{\widehat{\Pr}_{\mathbf{G}}(\boldsymbol{\gamma} \mid \tau, \mathbf{g})} \right|_{\boldsymbol{\gamma}=\boldsymbol{\gamma}^*(\tau)},
\]
where $\widehat{\Pr}_{\mathbf{G}}(\boldsymbol{\gamma} \mid \tau, \mathbf{g})$ is the Gaussian approximation generated from a Taylor expansion around $\boldsymbol{\gamma}^*(\tau)$, the mode of $\Pr(\boldsymbol{\gamma} \mid \tau, \mathbf{g})$ for a given $\tau$.
We can find $\boldsymbol{\gamma}^*(\tau)$ using the Newton-Raphson method. We next approximate the distribution of $\gamma_i$ conditional on $\tau$ with
\[
\widehat{\Pr}(\gamma_i \mid \tau, \mathbf{g}) \propto \left. \frac{\Pr(\boldsymbol{\gamma}, \tau, \mathbf{g})}{\widehat{\Pr}_{\mathbf{G}}(\boldsymbol{\gamma}_{-i} \mid \tau, \mathbf{g})} \right|_{\boldsymbol{\gamma}=\boldsymbol{\gamma}^*_{-i}},
\]
where $\widehat{\Pr}_{\mathbf{G}}(\boldsymbol{\gamma}_{-i} \mid \tau, \mathbf{g})$ and $\boldsymbol{\gamma}^*_{-i} = \E_G(\boldsymbol{\gamma}_{-i} \mid \gamma_i, \tau, \mathbf{g})$ are computed using $\widehat{\Pr}_{\mathbf{G}}(\boldsymbol{\gamma} \mid \tau, \mathbf{g})$. Finally, we normalize and combine the two approximations, then use numerical integration to calculate
\[
\widehat{\Pr}(\gamma_i \mid \mathbf{g}) = \int{\widehat{\Pr}(\gamma_i \mid \tau, \mathbf{g}) \widehat{\Pr}(\tau \mid \mathbf{g}) d\tau}.
\]

The outline for BNPR-PS is very similar.
The approximate marginal of the hyperparameters is 
\[
\widehat{\Pr}(\tau, \boldsymbol{\beta} \mid \mathbf{g}, \mathbf{s}) \propto \left. \frac{\Pr(\boldsymbol{\gamma}, \tau, \boldsymbol{\beta}, \mathbf{g}, \mathbf{s})}{\widehat{\Pr}_{\mathbf{G}}(\boldsymbol{\gamma} \mid \tau, \boldsymbol{\beta}, \mathbf{g}, \mathbf{s})} \right|_{\boldsymbol{\gamma}=\boldsymbol{\gamma}^*(\tau, \boldsymbol{\beta})},
\]
for similarly defined factors.
We take advantage of an INLA extension by \citet{martins2013bayesian} that allows for multiple likelihoods.
The approximate distribution of $\gamma_i$ conditional on $\tau, \boldsymbol{\beta}$ becomes
\[
\widehat{\Pr}(\gamma_i \mid \tau, \boldsymbol{\beta}, \mathbf{g}, \mathbf{s}) \propto \left. \frac{\Pr(\boldsymbol{\gamma}, \tau, \boldsymbol{\beta}, \mathbf{g}, \mathbf{s})}{\widehat{\Pr}_{\mathbf{G}}(\boldsymbol{\gamma}_{-i} \mid \tau, \boldsymbol{\beta}, \mathbf{g}, \mathbf{s})} \right|_{\boldsymbol{\gamma}=\boldsymbol{\gamma}^*_{-i}},
\]
and the final numerical integration is analogously more complex but still tractable, since we integrate over both $\tau$ and $\boldsymbol{\beta}$.

We use the R-INLA package \citep{rue2009approximate, martins2013bayesian} to perform
the above calculations.
We make INLA approximations of BNPR and BNPR-PS posteriors available, along with other phylodynamic tools, in the R package \texttt{phylodyn} which can be found at \href{https://github.com/mdkarcher/phylodyn}{https://github.com/mdkarcher/phylodyn}.

\section*{Simulation study}

We investigate estimating effective population size in the presence of preferential sampling via simulated data.
First, we seek to show where and how the model misspecification resulting from ignoring preferential sampling manifests itself
in terms of posterior median and Bayesian credible interval width estimation.
Our second goal is to show what we gain by properly modeling preferential sampling.

Our primary set of simulation results use the family of seasonally-varying effective population size functions characterized by
\begin{equation}
\label{eq:ne}
N_{e,a,o}(t) = \begin{cases} 10 + 90/(1+\exp\{a[3-(t+o \mbox{ (mod 12)})]\}), & \mbox{if } t+o \mbox{ (mod 12) } \leq 6, \\
10 + 90/(1+\exp\{a[3 + (t+o \mbox{ (mod 12)}) - 12]\}) , & \mbox{if } t+o \mbox{ (mod 12) } > 6. \end{cases}
\end{equation}
For all of our experiments, the smoothness parameter $a = 2$ will be used.
This family emulates a cyclical population time series with $t$ in nominal months.
The shape is loosely modeled after flu seasons, with $o$ controlling which part of the year $t=0$ represents ($o=0,3,6$ emulates summer, spring, and winter, respectively).
We simulate genealogies with varying tip sampling times using two sampling schedules.
The uniform schedule distributes $n$ sampling times uniformly throughout a given sampling interval.
The proportional schedule distributes sampling times in the sampling interval according to an inhomogeneous Poisson process with intensity proportional to effective population size.
The proportionality constant here is tuned to have an expected number of sampling times equal to $n$.

\begin{figure}[htbp]
	\centering
	\includegraphics[width=1.0\textwidth]{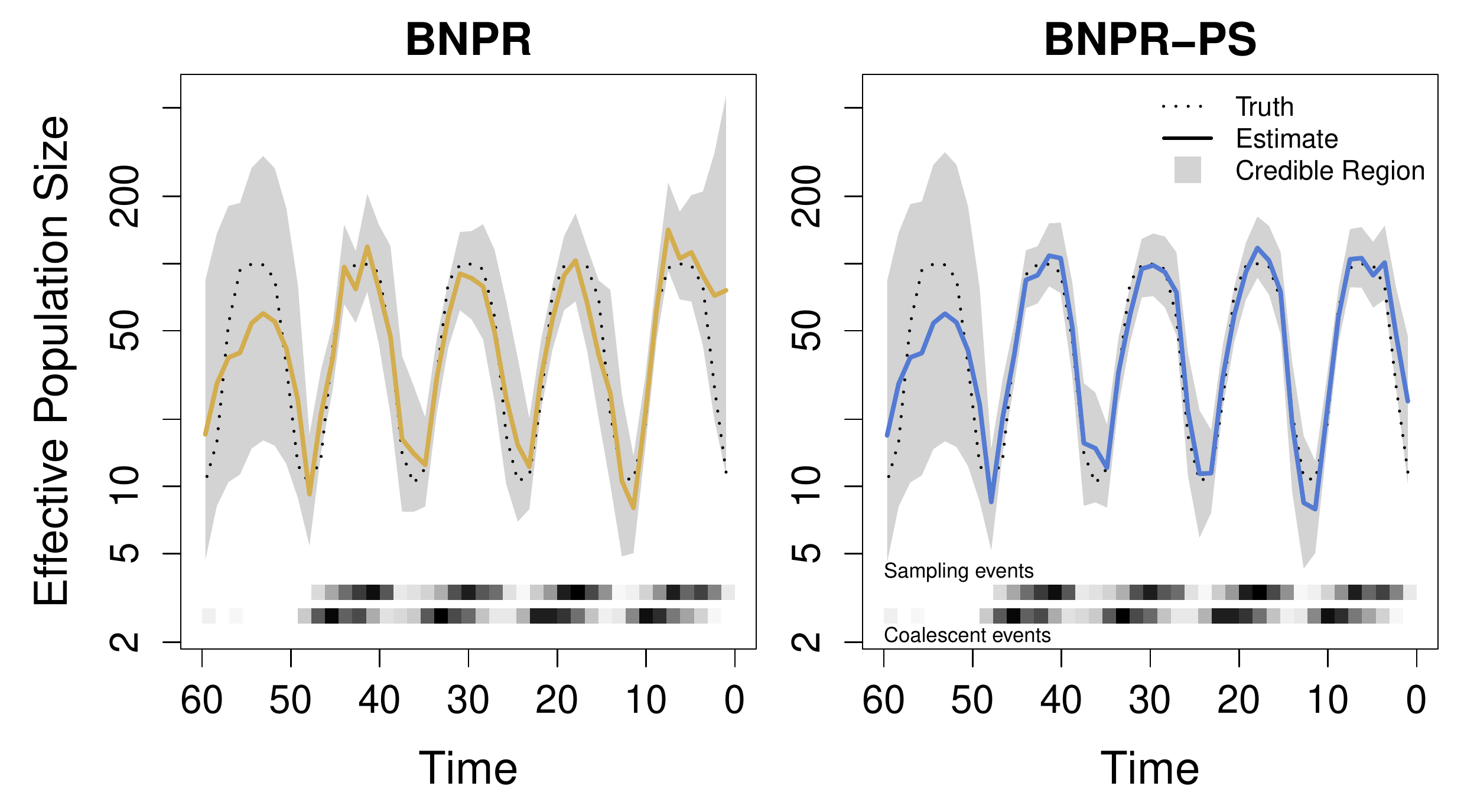}
	\caption{Graphical representation of the output of a single genealogy simulation and integrated nested Laplace approximation (INLA) estimation.
		The dotted black lines represent the true population trajectory.
		The solid colored lines represent the posterior median estimates, while the shaded regions represent the 95\% credible regions.
		At bottom, the upper and lower heatmaps represent frequencies of sampling events and coalescent events, respectively.
		For this figure, we sampled individuals according to an inhomogeneous Poisson process with intensity proportional to effective population size $N_e(t)$.
		The plot on the left is generated by Bayesian nonparametric phylodynamic reconstruction (BNPR) and does not account for preferential sampling, while the plot on the right is generated by Bayesian nonparametric phylodynamic reconstruction with preferential sampling (BNPR-PS) and incorporates our sampling time model.
		Time is in months.}
	\label{fig:Single}
\end{figure}

We explore the properties of our two methods using a Monte Carlo approach.
To create a Monte Carlo iteration, we generate our sampling times according to their sampling schedules, then simulate our genealogies using coalescent theory via the rejection sampling method of \citet{palacios2013gaussian}.
Given the genealogy and the samples, we infer the effective population time series, using BNPR and BNPR-PS to approximate grids of marginal posteriors.
For each iteration, this gives us approximate estimates of the posterior median and quantiles at each point in the effective population size time series.
In Figure \ref{fig:Single}, we see outputs from BNPR and BNPR-PS on the same example iteration.

Our first set of experiments is aimed at determining the extent of the bias introduced by unaccounted preferential sampling.
With $r$ Monte Carlo iterations, we take two approaches to locating model misspecification error---time interval analysis and pointwise analysis.
For time interval analyses, we calculate summary statistics for a pre-specified time interval $(a,b)$ and average them over the set of $r$ simulation iterations.
For pointwise analyses however, we consider the time series of point estimates from each iteration, and then on a pointwise basis we calculate aggregate point estimates and confidence intervals.

Our time interval summary statistics are
\textit{mean relative deviation},
\[
\text{MRD} = \frac{1}{r} \sum_{i=1}^r \left[ \frac{1}{b-a}\int_a^b \frac{|\hat{N}^{\boldsymbol{\gamma}}_{i}(t) - N^{\boldsymbol{\gamma}}(t)|}{N^{\boldsymbol{\gamma}}(t)} dt \right],
\]
\textit{mean relative width} of the 95\% Bayesian credible intervals,
\[
\text{MRW} = \frac{1}{r} \sum_{i=1}^r \left[ \frac{1}{b-a}\int_a^b
\frac{\hat{N}^{\boldsymbol{\gamma}}_{i,0.975}(t) -
\hat{N}^{\boldsymbol{\gamma}}_{i,0.025}(t)}{N^{\boldsymbol{\gamma}}(t)} dt \right] ,
\]
where $N^{\boldsymbol{\gamma}}(t)$ is the discretized true effective population size trajectory,
$\hat{N}^{\boldsymbol{\gamma}}_{i}(t)$ is the estimated posterior median of effective population sizes for iteration $i$,
and $\hat{N}^{\boldsymbol{\gamma}}_{i,q}(t)$ is the estimated $q$th posterior quantile for iteration $i$.
We also look at \textit{mean envelope}, ME, the proportion of grid points where the credible interval contains the true trajectory, averaged over all grid points contained in $[a,b]$ across all Monte Carlo iterations.

For a given grid of time points $\{t_j\}_{j=0}^k$, pointwise analysis computes the means of pointwise posterior medians,
\[
\text{mpmedian}(t_j) = \frac{1}{r} \sum_{i=1}^{r}{\hat{N}^{\boldsymbol{\gamma}}_{i,0.5}(t_j)}, \text{ for } j=0,\ldots,k,
\]
pointwise mean relative errors,
\[
\text{mre}(t_j) = \frac{1}{r} \sum_{i=1}^{r}{\frac{\hat{N}^{\boldsymbol{\gamma}}_{i,0.5}(t_j) - N^{\boldsymbol{\gamma}}(t_j)}{N^{\boldsymbol{\gamma}}(t_j)}}, \text{ for } j=0,\ldots,k,
\]
and a sequence of mean relative widths of the pointwise Bayesian credible intervals,
\[
\text{mrw}(t_j) = \frac{1}{r} \sum_{i=1}^{r}{\frac{\hat{N}^{\boldsymbol{\gamma}}_{i,0.975}(t_j) - \hat{N}^{\boldsymbol{\gamma}}_{i,0.025}(t_j)}{N^{\boldsymbol{\gamma}}(t_j)}}, \text{ for } j=0,\ldots,k.
\]
We choose grid size $k=300$, number of simulation iterations $r=512$, and expected number of lineages per genealogy $n=500$. We choose the sampling interval $[0,48]$ for all simulations.

\subsection*{Ignoring preferential sampling}
Table \ref{tab:SimRunwide} shows the averaged time interval summary statistics for simulated genealogies under uniform and proportional schedules for the time intervals $(0,6)$ and $(6,48)$.
Genealogies were simulated assuming effective population size function $N_{e,2,0}(t)$ defined in Equation \ref{eq:ne}.
We show the time interval summary statistics for inferred effective population sizes both ignoring and considering preferential sampling.
Ignoring preferential sampling (Table 1 under BNPR),
we note a 17\% increase in mean relative deviation from uniform to proportional schedules,
as well as a 20\% increase in mean relative width of Bayesian credible intervals for $(6,48)$.
For $(0,6)$ the increase is more stark.
We see a 407\% increase in mean relative deviation from uniform to proportional, and a 799\% increase in mean relative width of Bayesian credible intervals.
Under proportional sampling, we see a notable increase in mean envelope, ME, on the $(0,6)$ interval.
All other cases show BNPR and BNPR-PS having ME within Monte Carlo error.
These results confirm that ignoring preferential sampling affects both bias and variance of Bayesian nonparametric estimators of the effective populations size.

\begin{figure}[htbp]
	\centering
		\includegraphics[width=1.0\textwidth]{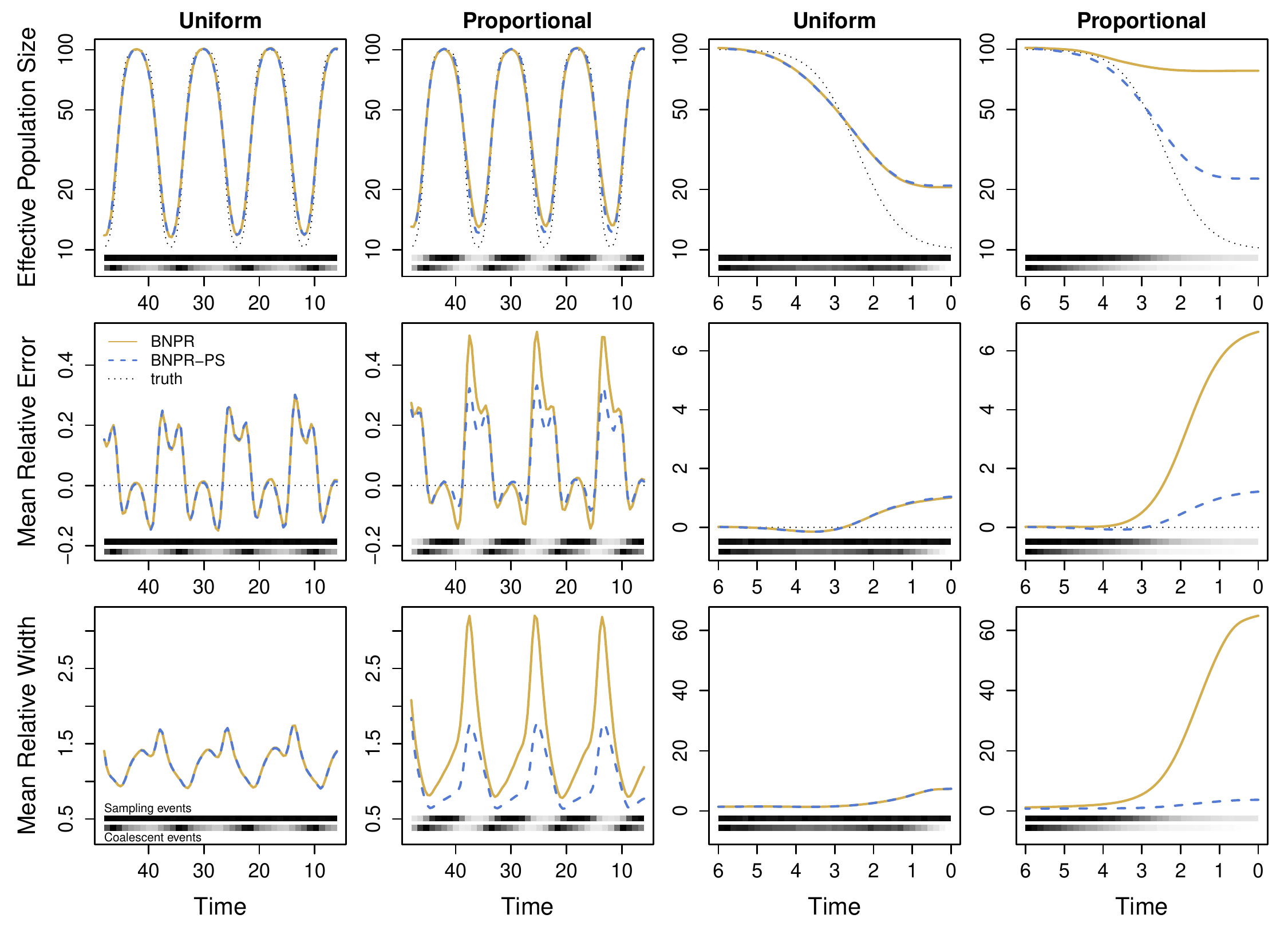}
	\caption{Comparison of pointwise statistics; dotted black lines represent the truth, where applicable.
		Solid yellow lines represent the conditional method BNPR (ignoring preferential sampling),
		while dashed blue lines represent the sampling-aware method BNPR-PS (accounting for preferential sampling).
		The first row shows true and estimated effective population sizes, the second shows mean relative error, while the third shows mean relative width of the 95\% Bayesian credible interval.
		The left two columns show the interval $(6,48)$ where both models perform at their best.
		The right two columns show $(0,6)$, where BNPR-PS performs significantly better.
		At the bottom of each plot, the distribution of sampling events (above) and coalescent events (below) are shown as heat maps.
		Time is in months.}
	\label{fig:Comparison}
\end{figure}

Figure \ref{fig:Comparison} (solid lines) compares the average pointwise statistics for the uniform and proportional sampling schedules.
Note the marked increase in mean relative error in several locations.
We also see much larger mean relative widths in the same locations.
We conjecture that these features are representative of the model misspecification error that we would expect while sampling sequences/lineages preferentially in time but not accounting for it in the model.

\begin{table}[htbp]
	\centering
		\begin{tabular}{lrrrrrrrr}
		\toprule
     & \multicolumn{2}{c}{Uniform---$(6,48)$} & \multicolumn{2}{c}{Proportional---$(6,48)$} & \multicolumn{2}{c}{Uniform---$(0,6)$} & \multicolumn{2}{c}{Proportional---$(0,6)$} \\
     \cmidrule(r){2-3}  \cmidrule(r){4-5} \cmidrule(r){6-7}  \cmidrule(r){8-9}
     &    BNPR & BNPR-PS &    BNPR & BNPR-PS &    BNPR & BNPR-PS &    BNPR & BNPR-PS \\
       \midrule
MRD & 0.205 & 0.205 & 0.239 & 0.183 & 0.430 & 0.436 & 2.181 & 0.432 \\
MRW & 1.255 & 1.255 & 1.500 & 1.008 & 2.816 & 2.816 & 19.681 & 1.682 \\
ME   & 0.965 & 0.964 & 0.962 & 0.957 & 0.950 & 0.948 & 0.833 & 0.898 \\
    \bottomrule
		\end{tabular}
		\caption{Averaged time interval summary statistics for BNPR and BNPR-PS.
			We compare the performance of the models under two different sampling distributions.
			Uniform distributes sampling times according to a uniform distribution on the interval $(0,48)$, while proportional distributes sampling times according to a inhomogeneous Poisson process with intensity proportional to effective population size $N_e(t)$ on the same interval.
			We examine the statistics mean relative deviation (MRD), mean relative width of the 95\% Bayesian credible interval (MRW), and mean envelope (ME).
			We average over statistics over the interval $(6,48)$ where both methods perform well and over the most recent interval $(0,6)$ where BNPR-PS performs considerably better.}
  \label{tab:SimRunwide}
\end{table}

\begin{figure}[htbp]
	\centering
		\includegraphics[width=1.0\textwidth]{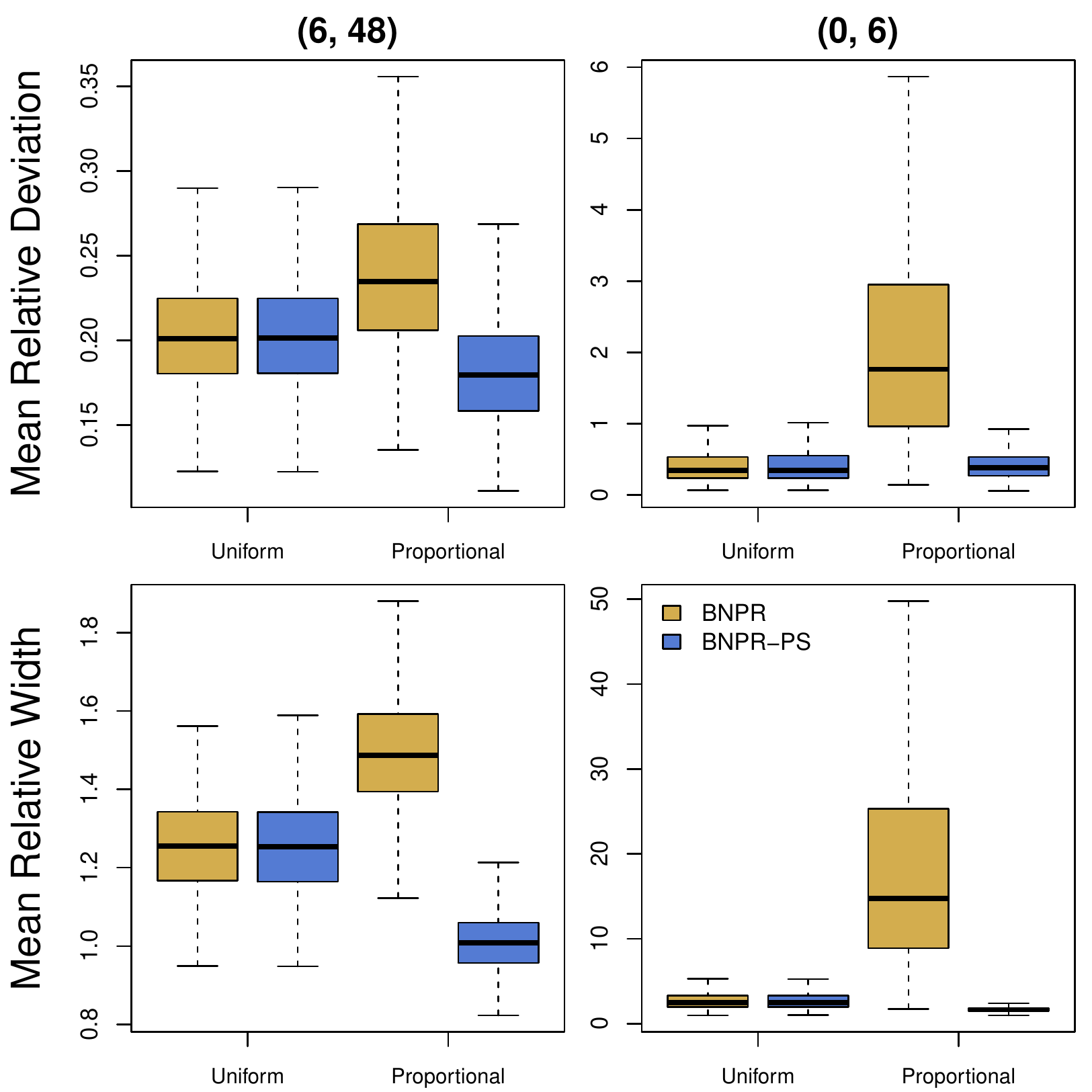}
	\caption{Comparison of time interval statistics.
		Within each plot, we apply BNPR and BNPR-PS to sampling times generated according to a Uniform distribution on the left and proportionally to effective population size on the right.
		In the left column of plots, we examine the interval $(6,48)$ where the performances of both models are comparable.
		In the right column, we show $(0,6)$, and note that BNPR-PS performs well, while BNPR performs considerably worse.}
	\label{fig:ComparisonBoxplots}
\end{figure}

\subsection*{Accounting for preferential sampling}

Table \ref{tab:SimRunwide} under BNPR-PS shows the time interval statistics for the sampling-aware model.
For interval $(6,48)$, mean relative deviation decreases by 23\% versus BNPR under proportional sampling,
while mean relative width of Bayesian credible intervals decreases by a larger margin of 33\%.
For interval $(0,6)$ mean relative deviation and mean relative width decrease by 80\% and 91\%, respectively.
Under uniform sampling, BNPR-PS performs almost identically to BNPR for both intervals.

Figure \ref{fig:Comparison} (dashed lines) compares the average pointwise statistics for the uniform and proportional sampling schedules under BNPR-PS.
We see that BNPR-PS does not experience the increase in relative error that BNPR experiences under preferential sampling.
The plots also show an improvement in mean relative width of Bayesian credible intervals under preferential sampling due to the additional information available.

\subsection*{Negative control simulations}

In the previous sections, we find a pattern of increased mean relative deviation and mean relative width  while using a conditional model in a scenario involving preferential sampling.
However, it is possible that this behavior of the conditional, state-of-the-art coalescent model can be seen under other simulation scenarios that
cluster sampling times, even when such clustering has no relationship to the effective population size fluctuations.
To test this assertion, we design a pair of negative control simulation studies to have random clusters of sampling times, but no preferential sampling.

First, we apply BNPR to genealogies generated from randomly constructed piecewise constant sampling intensity functions, independent of effective population size; see Figure \ref{fig:PCComparison}.
We see some examples of increased mean relative error, but nothing as consistent nor prevalent as in the preferential sampling case above (see Figures \ref{fig:Single} and \ref{fig:Comparison}).
Similarly, we see increased mean relative width in several locations, but decreased widths in others.
Second, we apply BNPR to genealogies generated from Gaussian process evaluations (subsampled for relatively similar shapes and number of peaks and troughs to the true population trajectory); see Figure \ref{fig:GPComparison}.
This model has shape characteristics closer to the population trajectory since we are sampling from a Gaussian process.
Despite the similar shapes, we see fewer increases in mean relative width and smaller increases in mean relative width.
We conclude that unaccounted preferential sampling produces markedly more error more consistently than the negative control cases.

We also apply BNPR-PS to the same scenarios as above (see Figure \ref{fig:PCComparison}).
BNPR-PS's performance suffers significantly due to both scenarios violating its fundamental assumption of a fixed relationship between effective population size and sampling intensity.
We see BNPR-PS performs worst locally when there is a nearly fixed relationship which is suddenly reversed in a small time interval.

\section*{Case studies}

\begin{table}[htbp]
	\centering
		\begin{tabular}{lcccrrrr}
		\toprule
    & & \multicolumn{2}{c}{EMRW} & \multicolumn{2}{c}{ 95\% credible} & \multicolumn{2}{c}{ 95\% credible} \\
     \cmidrule(r){3-4}
    & $n$ &    BNPR & BNPR-PS & \multicolumn{2}{c}{interval of $\beta_0$} & \multicolumn{2}{c}{interval of $\beta_1$}\\
       \midrule
\textbf{New York influenza} & 709  & 1.23 & 0.58 & (-47.4, & -30.3) & (5.88, & 10.23)\\
\midrule
\textbf{Regional influenza} & & & & & & & \\
 \midrule
USA \& Canada & 520 & 1.83 & 1.11 & (-3.02, & -0.79) & (2.52, & 4.05) \\
South America & 191 & 0.86 & 0.91 & (-4.21, & -0.42) & (3.27, & 7.52) \\
Europe & 361 & 1.73 & 0.96 & (-6.61, & -2.44) & (3.68, & 6.88) \\
India & 233 & 1.79 & 1.30 & (-2.18, & 0.50) & (2.34, & 4.78) \\
Japan \& Korea & 444 & 1.82 & 1.09 & (-2.23, & -0.25) & (2.35, & 3.76) \\
North China & 384 & 1.80 & 1.09 & (-2.63, & -0.27) & (2.22, & 3.89) \\
South China & 528 & 1.27 & 0.78 & (-1.05, & 1.00) & (1.68, & 3.23) \\
Southeast Asia & 494 & 0.99 & 0.54 & (-7.93, & -2.55) & (4.39, & 8.86) \\
Oceania & 461 & 1.53 & 0.88 & (-1.51, & 0.43) & (2.71, & 4.52) \\
    \bottomrule
		\end{tabular}
		\caption{Table of empirical mean relative widths and Bayesian credible intervals of $\beta_1$ for our real-world examples.
			The regional influenza dataset is broken down into world regions.
			In all but one region, we see improvements, or at worst near-equality, in empirical mean relative width (EMRW) using BNPR-PS over BNPR.
			}
  \label{tab:RealWorld}
\end{table}

\subsection*{New York influenza}

We base our first case study on a subset of the data from \citet{rambaut2008genomic}, also analyzed by \citet{palacios2013gaussian}.
We focus on the 709 hemagglutinin gene sequences of H3N2 human influenza type A obtained from the National Center for Biotechnology Information (NCBI) Influenza Virus Sequence Database for years 1992 through 2005 from New York State.
We align the sequences using the software MUSCLE \citep{MUSCLE}, and infer a maximum clade credibility genealogy using the software BEAST \citep{BEAST}.
We infer the genealogy branch lengths in units of years using a strict molecular clock,
a constant effective population size prior,
and an HKY substitution model with the first two nucleotides of a codon sharing the same estimated transition matrix,
while the third nucleotide's transition matrix is estimated separately.
We then apply our two algorithms to the estimated genealogy.

\begin{figure}[htbp]
	\centering
		\includegraphics[width=1.0\textwidth]{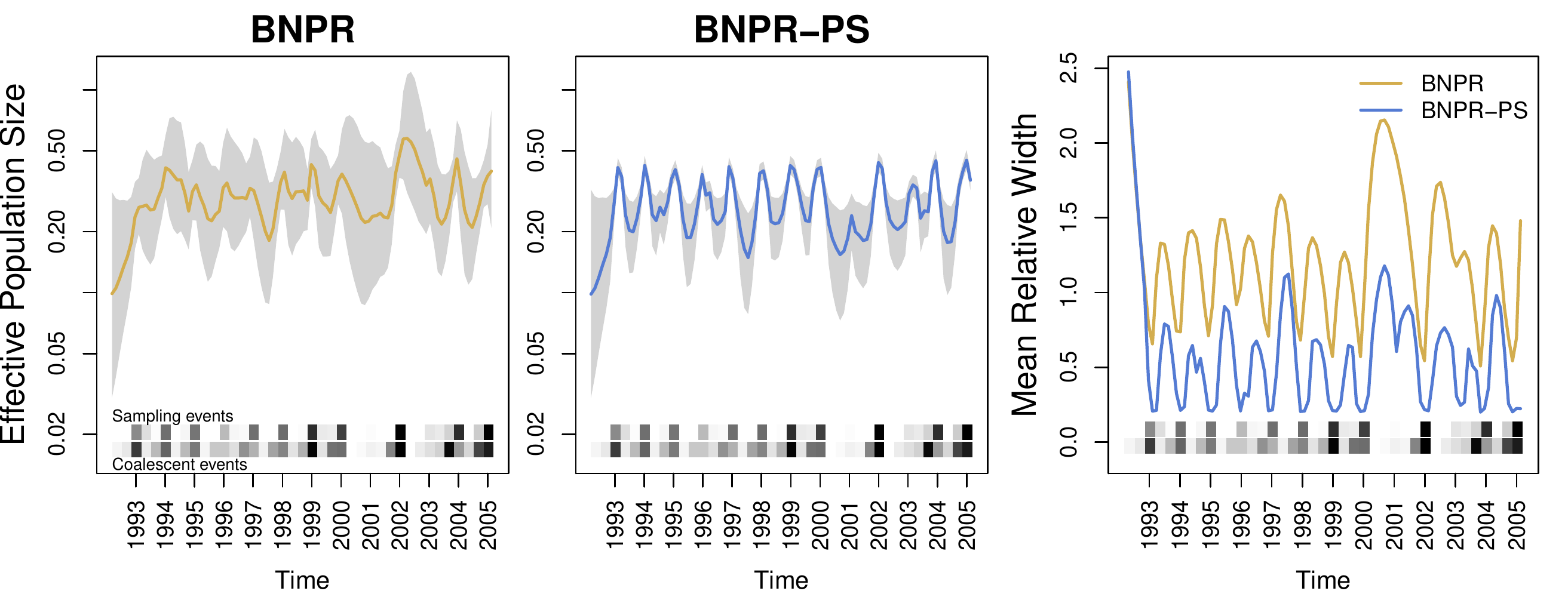}
	\caption{BNPR and BNPR-PS models applied to the genealogy inferred from
	the influenza example from \citep{rambaut2008genomic}.
	Years mark January of the corresponding year.
	Note the correlation of higher effective population size $N^{\boldsymbol{\gamma}}(t)$ with more intense sampling frequencies (darker regions in the Sampling events heatmap), suggesting preferential sampling.
	We see a marked improvement in discerning the seasonal influenza patterns and significantly thinner credible regions under BNPR-PS.}
	\label{fig:NewYorkInfluenza}
\end{figure}

We find that BNPR produces results in line with previous analyses of this dataset, showing a characteristic uncertainty around the flu seasons of 2000-2001 and 2002-2003 (see Figure \ref{fig:NewYorkInfluenza}).
In contrast, BNPR-PS shows a marked improvement in the regularity of the reconstructed flu seasons, as well as thinner Bayesian credible intervals across the the whole observation interval.
Estimations also improved during the
unusual flu seasons of 2000-2001 and 2002-2003, consistent with these seasons being H1N1 dominant seasons instead of H3N2 dominant \citep{goldstein2011predicting}.

To compare performance of the BNPR and BNPR-PS models, we introduce an empirical measure of performance because we cannot know the true population size trajectory.
We calculate the time interval and pointwise \textit{empirical mean relative width} (EMRW) of the 95\% Bayesian credible intervals,
\[
\text{EMRW} = \frac{1}{r} \sum_{i=1}^r \left[ \frac{1}{b-a}\int_a^b \frac{\hat{N}^{\boldsymbol{\gamma}}_{i,0.975}(t) - \hat{N}^{\boldsymbol{\gamma}}_{i,0.025}(t)}{\hat{N}^{\boldsymbol{\gamma}}_{i}(t)} dt \right], \text{ for } j=0,\ldots,k,
\]
and
\[
emrw(t_j) = \frac{1}{r} \sum_{i=1}^{r}{\frac{\hat{N}^{\boldsymbol{\gamma}}_{i,0.975}(t_j) - \hat{N}^{\boldsymbol{\gamma}}_{i,0.025}(t_j)}{\hat{N}^{\boldsymbol{\gamma}}_{i}(t_j)}}, \text{ for } j=0,\ldots,k.
\]
Table \ref{tab:RealWorld} shows a very high value of $\beta_1$ for this dataset, suggesting a strong pattern of preferential sampling, and accordingly we see a marked improvement of BNPR-PS model over its BNPR counterpart in estimation precision as measured by the Bayesian credible interval widths (EMRW).

\subsection*{Regional influenza}

\citet{zinder2014seasonality} examine world-wide seasonal patterns of migration of H3N2 influenza across the regions of the world.
They also examine different seasonal incidence patterns, with tropical regions having a relatively flat incident rate throughout the year, while temperate regions show larger seasonal variation with higher incidence in winter months.
In order to explore the effects of seasonality on preferential sampling, we examine the regions separately.
We align the sequences using the software MUSCLE \citep{MUSCLE}, and infer a maximum clade credibility genealogy using the software BEAST \citep{BEAST}.
We infer the genealogy branch lengths in units of years using a strict molecular clock,
a constant effective population size prior,
and an HKY substitution model with the first two nucleotides of a codon sharing the same estimated transition matrix,
while the third nucleotide's transition matrix is estimated separately.
We then apply our two algorithms to the estimated genealogy.

We find that none of the regions contain 0 in their $\beta_1$ Bayesian credible interval (see Table \ref{tab:RealWorld}),
suggesting a relationship between effective population size and sampling frequency.
Across all regions except South America, we see improvements of the BNPR-PS model over the BNPR model in estimation precision (EMRW).
We examine three of the regions more closely in Figures \ref{fig:RegionalInfluenza} and \ref{fig:RegionalSeasonality} and the remaining six regions in the appendix.
We see noticeable improvements in the relative widths of the Bayesian credible intervals.
We also see more pronounced seasonality in the estimated effective population size trajectories produced by BNPR-PS.
The USA/Canada region shows the expected seasonal peak in January-February,
while the Oceania region shows the same in July-September.
South China shows less seasonality overall, but BNPR-PS shows a more pronounced August peak despite the region being in the northern hemisphere.
This is, however, in line with previous findings, most likely due to southern China's more tropical climate \citep{shu2010dual}.

\begin{figure}[htbp]
	\centering
		\includegraphics[width=1.0\textwidth]{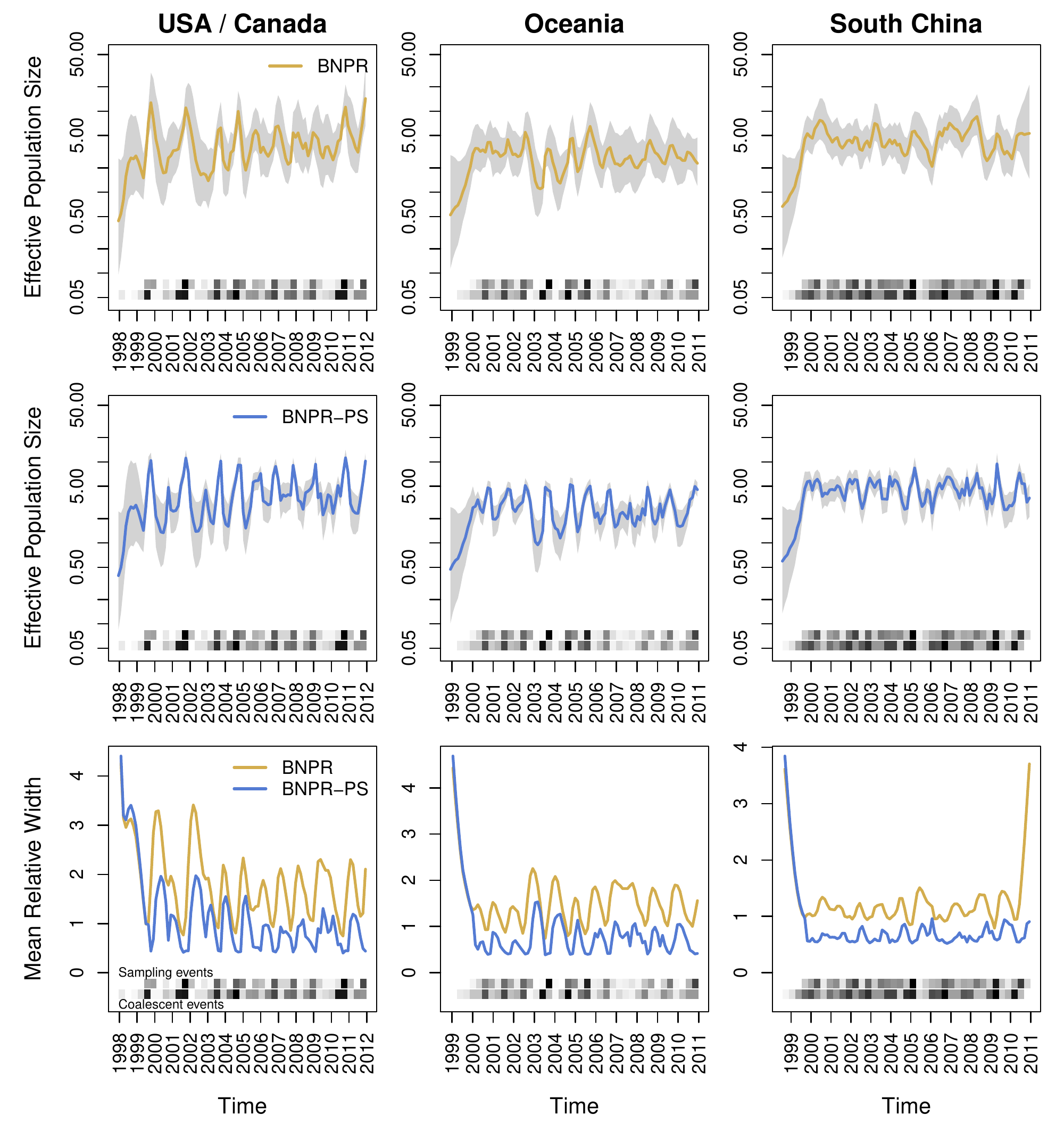}
	\caption{BNPR and BNPR-PS models applied to the genealogies inferred from the regional influenza example. We see moderate correlation between effective population size $N^{\boldsymbol{\gamma}}(t)$ and sampling frequencies in the data (Table \ref{tab:RealWorld}). We see improvements in Bayesian credible interval widths, and BNPR-PS performs as well or better than BNPR everywhere in these examples.}
	\label{fig:RegionalInfluenza}
\end{figure}

\begin{figure}[htbp]
	\centering
		\includegraphics[width=1.0\textwidth]{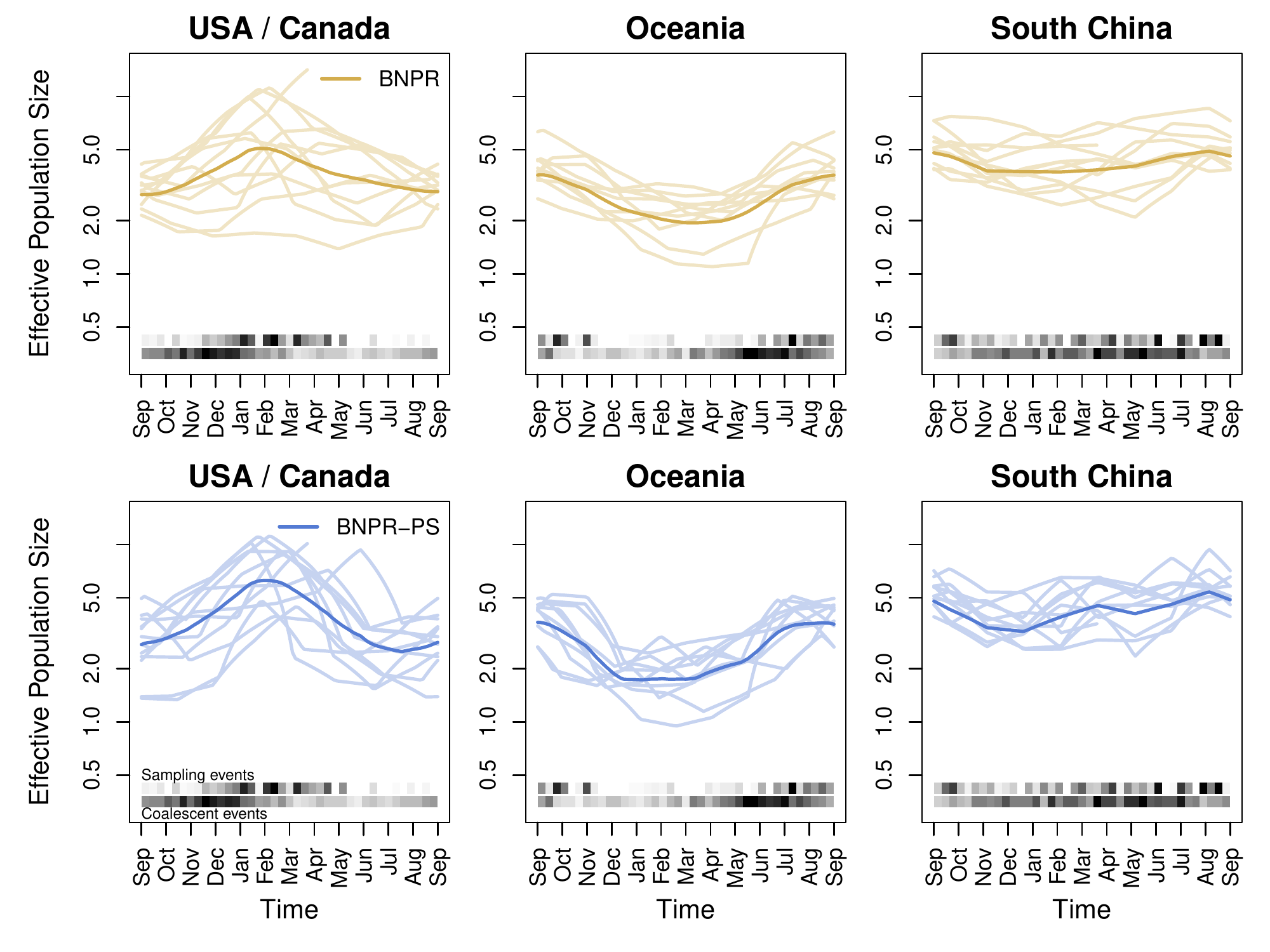}
	\caption{BNPR and BNPR-PS models applied to the genealogies inferred from the regional influenza example with years overlaid. Light lines correspond to year-specific estimates of the effective population size. The darker line in each plot is a pointwise average of light curves. We see more pronounced seasonality in the BNPR-PS plots.}
	\label{fig:RegionalSeasonality}
\end{figure}

\section*{Discussion}
Researchers who study measurably evolving populations \citep{drummond2003measurably}, such as viruses,
can inadvertently or purposefully preferentially select sequences in accordance to the changes in size of the population of interest.
Failing to account for such an ascertainment bias can compromise the statistical properties of phylodynamic inference.
Our simulation study shows that the effect of preferential sampling is particularly severe when the effective population size is decreasing.
We propose an extension to the state-of-the-art in Gaussian process-based Bayesian phylodynamic methods,
in which we assume that sampling times \textit{a priori} follow an inhomogeneous Poisson process with intensity proportional to a power of the effective population size.
This model extension eliminates the systematic estimation bias resulting from having unrecognized preferential sampling,
and also gives us better population size estimates by incorporating sampling times as an additional source of information.

Applied to the real-world examples, our method produces improvements over the state-of-the-art.
We see significantly improved precision, as well as more realistic estimation of seasonal variation of influenza diversity.
In the presence of weaker preferential sampling, as in some of the regional influenza examples,
we note that our method still performs better than the current state-of-the-art,
with no loss of performance aside from a slightly longer computation time.
In addition, by estimating $\beta_1$, the effect of population size on the log-intensity of sampling times,
we gain the ability to quantify the strength of the preferential sampling relationship in the different regions.
Such quantification is scientifically useful in infectious disease phylodynamics, because researchers may want to know whether frequency of sampling times can be used as a proxy for incidence.

One avenue of future exploration is to intentionally guarantee preferential sampling during the sequence data collection phase.
For example, if an epidemiological study contains noisy incidence data,
we can subsample sequences with intensity proportional to incidence and apply our sampling-aware BNPR-PS model to the resulting sequence data.
Such a procedure will indirectly combine sequence and incidence data to estimate the effective number of infections---a nontrivial task for the current methods \citep{Rasmussen2011}.
We contrast this to the approach of \citet{stack2010protocols}, who examined the effect of sampling infectious disease agent sequences in batches at different points in an epidemic's life-cycle compared to uniform and preferential sampling.
They found that their estimates were most accurate in terms of sum of squared differences when they sampled during a decline of the epidemics.
This is consistent with our results, as we see the most error and widest credible intervals during effective population size declines.
However, \citet{stack2010protocols} did not consider the effect of the relationship between their proposed sampling intensity and population size trajectories on estimation of population dynamics --- the primary goal of our work.

Our current implementation of the BNPR-PS model assumes a fixed, known genealogy.
However, in practice, genealogies are inferred with inherent uncertainty from sequence data.
One limitation of our method is that the INLA framework cannot be extended to include inference of genealogies.
However, it should be straightforward to incorporate the core of our approach---the sampling times model---into an MCMC sampler that targets the joint posterior distribution of population size trajectory,
genealogy of sampled sequences, and other parameters.
We intend to implement such an MCMC approach in the software BEAST \citep{BEAST}.

The main goal of this manuscript is to point out the danger of ignoring preferential sampling in phylodynamics.
Providing a solution to this problem, in the form of BNPR-PS model, remains our secondary goal,
but we emphasize that much work is still needed to refine our proposed approach.
The main weakness of our new model lies in its rigid parametric form of dependence between effective population size $N_e(t)$ and sequence sampling intensity $\lambda(t)$.
In our negative control simulations we see that BNPR-PS performance suffers, possibly greatly,
when this assumption of a fixed relationship between effective population size $N_e(t)$ and sampling intensity $\lambda(t)$ is violated.
Such model misspecification is most likely to occur if other variables besides effective population size $N_e(t)$ effect changes in the sampling intensity $\lambda(t)$.
For instance, not accounting for a lag between $N_e(t)$ and $\lambda(t)$ may cause a severe model misspecification.
Similarly, not accounting for increases in sampling intensity on longer time scales due to decreases in the cost of sequencing will bias our BNPR-PS estimation.
We plan to address these issues by modeling our sampling intensity $\lambda(t)$ as a log-linear combination of effective sample size and other covariates:
\[
\log[\lambda(t)] = \boldsymbol{\beta}^T \mathbf{c}(t),
\]
where $\mathbf{c}(t)^T = (1, N_e(t), c_1(t), \dots, c_p(t))$ and $c_i(t)$, $i=1,\dots,p$ are covariates of interest.
For example, the cost of genome sequencing over time and lagged population size $N_e(t-l)$
are among prime candidates for covariates to be included into our BNPR-PS model.
We hope to explore these model extensions in our future research.

\section*{Acknowledgments}
J.A.P.~acknowledges scholarship from CONACyT Mexico to pursue her research.
M.K., M.A.S, and V.N.M.~are supported by the NIH grant R01 AI107034.
M.A.S.~is supported by the NSF grant DMS 1264153 and NIH grant R01 LM012080.
V.N.M.~and T.B.~are supported by the NIH grant U54 GM111274.

\FloatBarrier

\bibliography{../bib/statgen}

\begin{thebibliography}{28}
\providecommand{\natexlab}[1]{#1}
\providecommand{\url}[1]{\texttt{#1}}
\expandafter\ifx\csname urlstyle\endcsname\relax
  \providecommand{\doi}[1]{doi: #1}\else
  \providecommand{\doi}{doi: \begingroup \urlstyle{rm}\Url}\fi

\bibitem[Diggle et~al.(2010)Diggle, Menezes, and Su]{diggle2010geostatistical}
P.~J. Diggle, R.~Menezes, and T.~Su.
\newblock Geostatistical inference under preferential sampling.
\newblock \emph{Journal of the Royal Statistical Society: Series C (Applied
  Statistics)}, 59\penalty0 (2):\penalty0 191--232, 2010.

\bibitem[Drummond et~al.(2002)Drummond, Nicholls, Rodrigo, and
  Solomon]{drummond2002estimating}
A.~J. Drummond, G.~K. Nicholls, A.~G. Rodrigo, and W.~Solomon.
\newblock Estimating mutation parameters, population history and genealogy
  simultaneously from temporally spaced sequence data.
\newblock \emph{Genetics}, 161\penalty0 (3):\penalty0 1307--1320, 2002.

\bibitem[Drummond et~al.(2005)Drummond, Rambaut, Shapiro, and
  Pybus]{drummond2005bayesian}
A.~J. Drummond, A.~Rambaut, B.~Shapiro, and O.~G. Pybus.
\newblock Bayesian coalescent inference of past population dynamics from
  molecular sequences.
\newblock \emph{Molecular Biology and Evolution}, 22\penalty0 (5):\penalty0
  1185--1192, 2005.

\bibitem[Drummond et~al.(2012)Drummond, Suchard, Xie, and Rambaut]{BEAST}
A.~J. Drummond, M.~A. Suchard, D.~Xie, and A.~Rambaut.
\newblock {B}ayesian phylogenetics with {BEAU}ti and the {BEAST} 1.7.
\newblock \emph{Molecular Biology and Evolution}, 29:\penalty0 1969--1973,
  2012.

\bibitem[Drummond et~al.(2003)Drummond, Pybus, Rambaut, Forsberg, and
  Rodrigo]{drummond2003measurably}
A.J. Drummond, O.G. Pybus, A.~Rambaut, R.~Forsberg, and A.G. Rodrigo.
\newblock Measurably evolving populations.
\newblock \emph{Trends in Ecology \& Evolution}, 18\penalty0 (9):\penalty0
  481--488, 2003.

\bibitem[Edgar(2004)]{MUSCLE}
R.~C. Edgar.
\newblock {MUSCLE}: {M}ultiple sequence alignment with high accuracy and high
  throughput.
\newblock \emph{Nucleic Acids Research}, 32:\penalty0 1792--1797, 2004.

\bibitem[Felsenstein and Rodrigo(1999)]{joseph_coalescent_1999}
J.~Felsenstein and A.~G. Rodrigo.
\newblock {C}oalescent {A}pproaches to {HIV} {P}opulation {G}enetics.
\newblock In \emph{The {E}volution of {HIV}}, pages 233--272. {Johns Hopkins
  University} Press, 1999.
\newblock ISBN 9780801861512.

\bibitem[Frost and Volz(2010)]{frost2010viral}
S.~D.~W. Frost and E.~M. Volz.
\newblock Viral phylodynamics and the search for an `effective number of
  infections'.
\newblock \emph{Philosophical Transactions of the Royal Society B: Biological
  Sciences}, 365\penalty0 (1548):\penalty0 1879--1890, 2010.

\bibitem[Gill et~al.(2013)Gill, Lemey, Faria, Rambaut, Shapiro, and
  Suchard]{skygrid}
M.~S. Gill, P.~Lemey, N.~R. Faria, A.~Rambaut, B.~Shapiro, and M.~A. Suchard.
\newblock Improving {B}ayesian population dynamics inference: a
  coalescent-based model for multiple loci.
\newblock \emph{Molecular biology and evolution}, 30\penalty0 (3):\penalty0
  713--724, 2013.

\bibitem[Goldstein et~al.(2011)Goldstein, Cobey, Takahashi, Miller, and
  Lipsitch]{goldstein2011predicting}
E.~Goldstein, S.~Cobey, S.~Takahashi, J.~C. Miller, and M.~Lipsitch.
\newblock Predicting the epidemic sizes of influenza {A/H1N1}, {A/H3N2}, and
  {B}: a statistical method.
\newblock \emph{PLoS medicine}, 8\penalty0 (7):\penalty0 952, 2011.

\bibitem[Grenfell et~al.(2004)Grenfell, Pybus, Gog, Wood, Daly, Mumford, and
  Holmes]{grenfell2004unifying}
B.~T. Grenfell, O.~G. Pybus, J.~R. Gog, J.~L.~N. Wood, J.~M. Daly, J.~A.
  Mumford, and E.~C. Holmes.
\newblock Unifying the epidemiological and evolutionary dynamics of pathogens.
\newblock \emph{Science}, 303\penalty0 (5656):\penalty0 327--332, 2004.

\bibitem[Griffiths and Tavar{\'e}(1994)]{griffiths1994sampling}
R.~C. Griffiths and S.~Tavar{\'e}.
\newblock Sampling theory for neutral alleles in a varying environment.
\newblock \emph{Philosophical Transactions of the Royal Society of London.
  Series B: Biological Sciences}, 344\penalty0 (1310):\penalty0 403--410, 1994.

\bibitem[Ho and Shapiro(2011)]{ho2011skyline}
S.~Y.~W. Ho and B.~Shapiro.
\newblock Skyline-plot methods for estimating demographic history from
  nucleotide sequences.
\newblock \emph{Molecular Ecology Resources}, 11\penalty0 (3):\penalty0
  423--434, 2011.

\bibitem[Holmes and Grenfell(2009)]{holmes2009discovering}
E.~C. Holmes and B.~T. Grenfell.
\newblock Discovering the phylodynamics of {RNA} viruses.
\newblock \emph{PLoS Computational Biology}, 5\penalty0 (10):\penalty0
  e1000505, 2009.

\bibitem[Kingman(1982)]{coalescent}
J.~F.~C. Kingman.
\newblock The coalescent.
\newblock \emph{Stochastic processes and their applications}, 13\penalty0
  (3):\penalty0 235--248, 1982.

\bibitem[Kuhner et~al.(1998)Kuhner, Yamato, and Felsenstein]{kuhner1998maximum}
M.~K. Kuhner, J.~Yamato, and J.~Felsenstein.
\newblock Maximum likelihood estimation of population growth rates based on the
  coalescent.
\newblock \emph{Genetics}, 149\penalty0 (1):\penalty0 429--434, 1998.

\bibitem[Martins et~al.(2013)Martins, Simpson, Lindgren, and
  Rue]{martins2013bayesian}
T.~G. Martins, D.~Simpson, F.~Lindgren, and H.~Rue.
\newblock Bayesian computing with {INLA}: new features.
\newblock \emph{Computational Statistics \& Data Analysis}, 67:\penalty0
  68--83, 2013.

\bibitem[Minin et~al.(2008)Minin, Bloomquist, and Suchard]{skyride}
V.~N. Minin, E.~W. Bloomquist, and M.~A. Suchard.
\newblock Smooth skyride through a rough skyline: {B}ayesian coalescent-based
  inference of population dynamics.
\newblock \emph{Molecular Biology and Evolution}, 25\penalty0 (7):\penalty0
  1459--1471, 2008.

\bibitem[Palacios and Minin(2012)]{palacios2012INLA}
J.~A. Palacios and V.~N. Minin.
\newblock Integrated nested {L}aplace approximation for {B}ayesian
  nonparametric phylodynamics.
\newblock In \emph{Proceedings of the Twenty-Eighth International Conference on
  Uncertainty in Artificial Intelligence}, pages 726--735, 2012.

\bibitem[Palacios and Minin(2013)]{palacios2013gaussian}
J.~A. Palacios and V.~N. Minin.
\newblock Gaussian process-based {B}ayesian nonparametric inference of
  population size trajectories from gene genealogies.
\newblock \emph{Biometrics}, 69\penalty0 (1):\penalty0 8--18, 2013.

\bibitem[Pybus et~al.(2000)Pybus, Rambaut, and Harvey]{pybus2000integrated}
O.~G. Pybus, A.~Rambaut, and P.~H. Harvey.
\newblock An integrated framework for the inference of viral population history
  from reconstructed genealogies.
\newblock \emph{Genetics}, 155\penalty0 (3):\penalty0 1429--1437, 2000.

\bibitem[Rambaut et~al.(2008)Rambaut, Pybus, Nelson, Viboud, Taubenberger, and
  Holmes]{rambaut2008genomic}
A.~Rambaut, O.~G. Pybus, M.~I. Nelson, C.~Viboud, J.~K. Taubenberger, and E.~C.
  Holmes.
\newblock The genomic and epidemiological dynamics of human influenza {A}
  virus.
\newblock \emph{Nature}, 453\penalty0 (7195):\penalty0 615--619, 2008.

\bibitem[Rasmussen et~al.(2011)Rasmussen, Ratmann, and Koelle]{Rasmussen2011}
D.A. Rasmussen, O.~Ratmann, and K.~Koelle.
\newblock Inference for nonlinear epidemiological models using genealogies and
  time series.
\newblock \emph{PLoS Computational Biology}, 7:\penalty0 e1002136, 2011.

\bibitem[Rue et~al.(2009)Rue, Martino, and Chopin]{rue2009approximate}
H.~Rue, S.~Martino, and N.~Chopin.
\newblock Approximate {B}ayesian inference for latent {G}aussian models by
  using integrated nested {L}aplace approximations.
\newblock \emph{Journal of the Royal Statistical Society: Series B},
  71\penalty0 (2):\penalty0 319--392, 2009.

\bibitem[Shu et~al.(2010)Shu, Fang, de~Vlas, Gao, Richardus, and
  Cao]{shu2010dual}
Yue-Long Shu, Li-Qun Fang, Sake~J de~Vlas, Yan Gao, Jan~Hendrik Richardus, and
  Wu-Chun Cao.
\newblock Dual seasonal patterns for influenza, china.
\newblock \emph{Emerging Infectious Diseases}, 16\penalty0 (4):\penalty0 725,
  2010.

\bibitem[Stack et~al.(2010)Stack, Welch, Ferrari, Shapiro, and
  Grenfell]{stack2010protocols}
J.~C. Stack, J.~D. Welch, M.~J. Ferrari, B.~U. Shapiro, and B.~T. Grenfell.
\newblock Protocols for sampling viral sequences to study epidemic dynamics.
\newblock \emph{Journal of The Royal Society Interface}, 7\penalty0
  (48):\penalty0 1119--1127, 2010.

\bibitem[Wakeley and Sargsyan(2009)]{wakeley2009extensions}
J.~Wakeley and O.~Sargsyan.
\newblock Extensions of the coalescent effective population size.
\newblock \emph{Genetics}, 181\penalty0 (1):\penalty0 341--345, 2009.

\bibitem[Zinder et~al.(2014)Zinder, Bedford, Baskerville, Woods, Roy, and
  Pascual]{zinder2014seasonality}
D.~Zinder, T.~Bedford, E.~B. Baskerville, R.~J. Woods, M.~Roy, and M.~Pascual.
\newblock Seasonality in the migration and establishment of {H3N2} influenza
  lineages with epidemic growth and decline.
\newblock \emph{BMC Evolutionary Biology}, 14\penalty0 (1):\penalty0 272, 2014.

\end{thebibliography}

\clearpage


\renewcommand{\thefigure}{A-\arabic{figure}}
\setcounter{figure}{0}

\renewcommand{\thetable}{A-\arabic{table}}
\setcounter{table}{0}

\renewcommand{\figurename}{Fig.}

\section*{Appendix}
\label{sec:AppendixA}

\subsection*{Hyperproportional simulations}

\begin{figure}[htbp]
	\centering
	\includegraphics[width=1.0\textwidth]{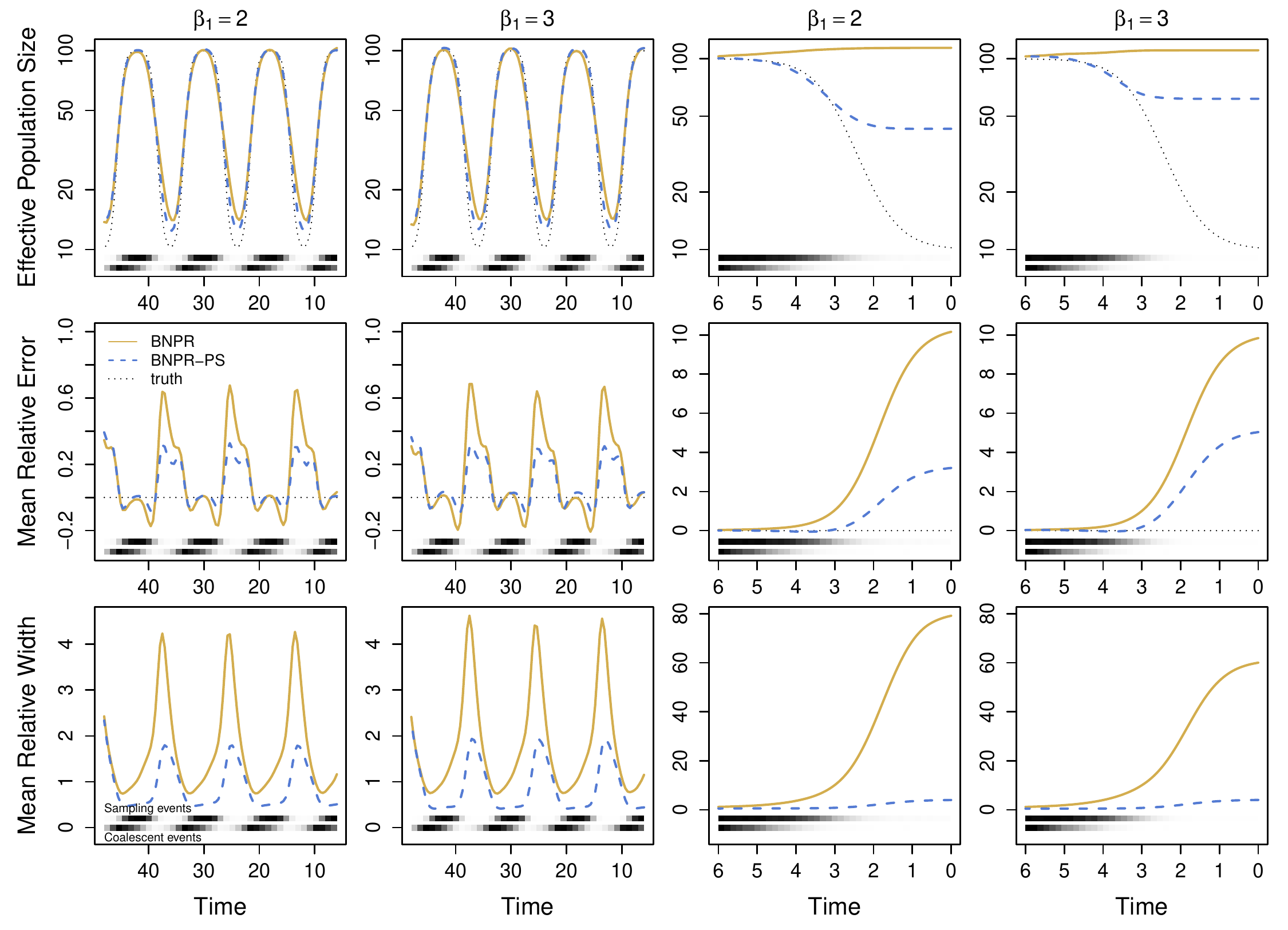}
	\caption{\small Comparison of pointwise statistics with hyperproportional preferential sampling; dotted lines represent the truth, where applicable.
		Solid yellow lines represent the conditional method BNPR (ignoring preferential sampling),
		while dashed blue lines represent the sampling-aware method BNPR-PS (accounting for preferential sampling).
		The first row shows true and estimated effective population sizes, the second shows mean relative error, while the third shows mean relative width of the 95\% Bayesian credible interval.
		The left two columns show the interval $(6,48)$ where both models perform at their best.
		The right two columns show $(0,6)$, where BNPR-PS performs significantly better.
		At the bottom of each plot, the distribution of sampling events (above) and coalescent events (below) are shown.
		Time is in months.}
	\label{fig:HighBetaComparison}
\end{figure}

To explore preferential sampling relationships with greater clustering than direct proportionality,
or \textit{hyperproportional} preferential sampling, we also perform a simulation study with $\beta_1 = 2,3$.
Fig. \ref{fig:HighBetaComparison} shows the pointwise statistics, while Table \ref{tab:HighBetaSimRunwide} lists the time interval statistics.
The results are largely consistent with $\beta_1 = 1$, but with slightly more bias under the BNPR.

\begin{table}[htbp]
	\centering
	\begin{tabular}{lrrrrrrrr}
		\toprule
		& \multicolumn{2}{c}{$\beta_1 = 2.0$ $(6,48)$} & \multicolumn{2}{c}{$\beta_1 = 3.0$ $(6,48)$} & \multicolumn{2}{c}{$\beta_1 = 2.0$ $(0,6)$} & \multicolumn{2}{c}{$\beta_1 = 3.0$ $(0,6)$} \\
		\cmidrule(r){2-3}  \cmidrule(r){4-5} \cmidrule(r){6-7}  \cmidrule(r){8-9}
		&    BNPR & BNPR-PS &    BNPR & BNPR-PS &    BNPR & BNPR-PS &    BNPR & BNPR-PS \\
		\midrule
		MRD  &   0.240 &   0.146 &   0.242 &   0.147 &    3.784 &   0.237 &   3.839 &   0.311  \\
		MRW  &   1.865 &   0.979 &   1.919 &   0.938 &  51.507 &   1.413 &   57.701 &   1.444  \\
		ME   &   0.989  &   0.987  &   0.990 &   0.978 &   0.852  &   0.988 &   0.891  &  0.972  \\
		\bottomrule
	\end{tabular}
	\caption{\small Averaged time interval summary statistics of the hyperproportional simulations over the interval $(6,48)$ where both methods perform well, and the most recent interval $(0,6)$ where BNPR-PS performs considerably better.}
	\label{tab:HighBetaSimRunwide}
\end{table}

\subsection*{Negative control simulations}

\begin{figure}[htb]
	\centering
		\includegraphics[width=1.0\textwidth, angle=0]{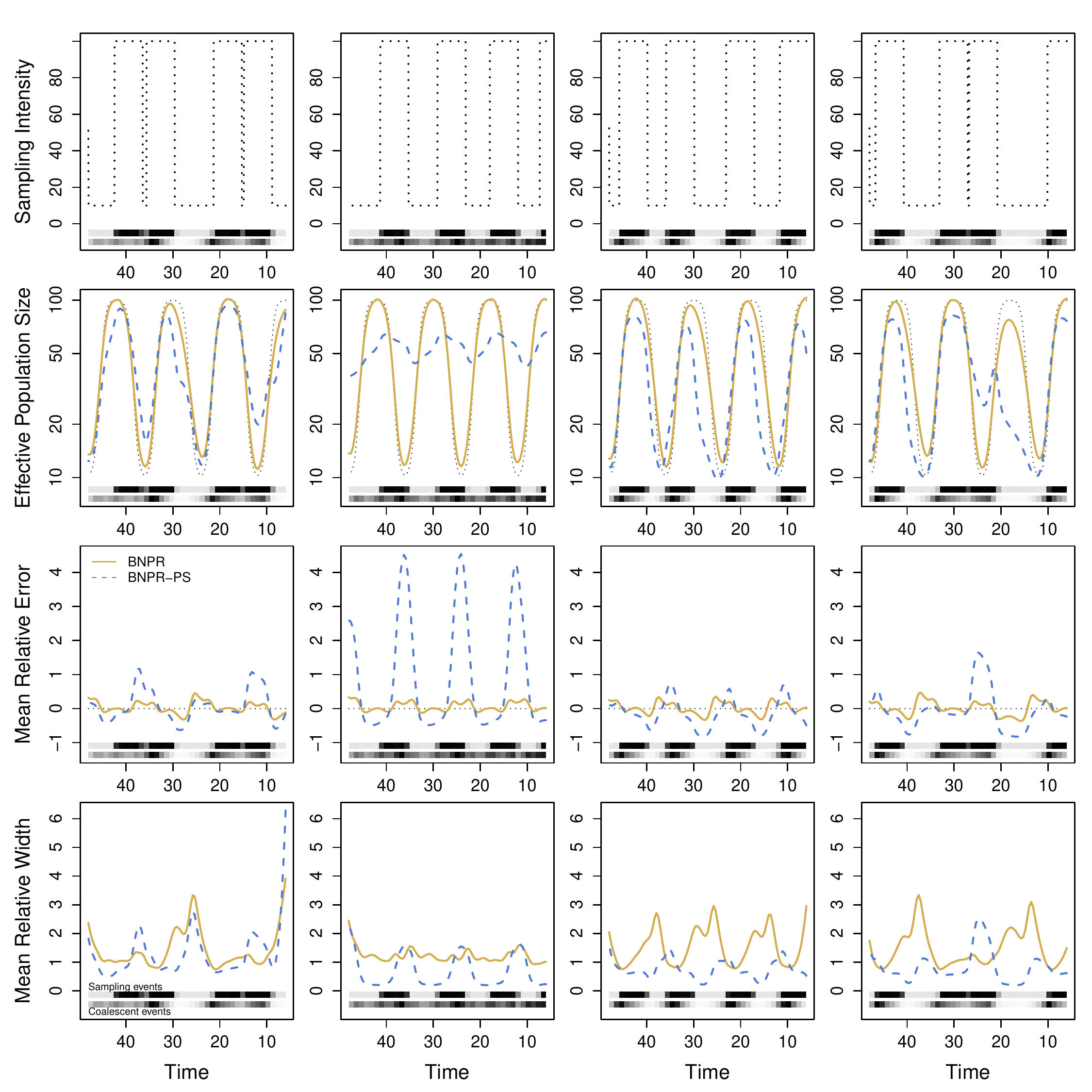}
	\caption{\small Comparison of pointwise statistics for a randomly generated piecewise constant sampling intensity trajectory independent of effective population size.
		Dotted lines represent the sampling intensity trajectory and true effective population size trajectory.
		Solid yellow lines represent the conditional method BNPR, while the dashed blue lines represent the sampling-aware model BNPR-PS.
		The first row shows the sampling intensity, the second shows true and estimated effective population sizes, the third shows mean relative error, while the fourth shows mean relative width of the 95\% Bayesian credible interval.
		The columns represent four realizations of the random sampling intensity trajectory.
		At the bottom of each plot, the distribution of sampling events (above) and coalescent events (below) are shown.}
	\label{fig:PCComparison}
\end{figure}

\begin{figure}[htb]
	\centering
		\includegraphics[width=1.0\textwidth]{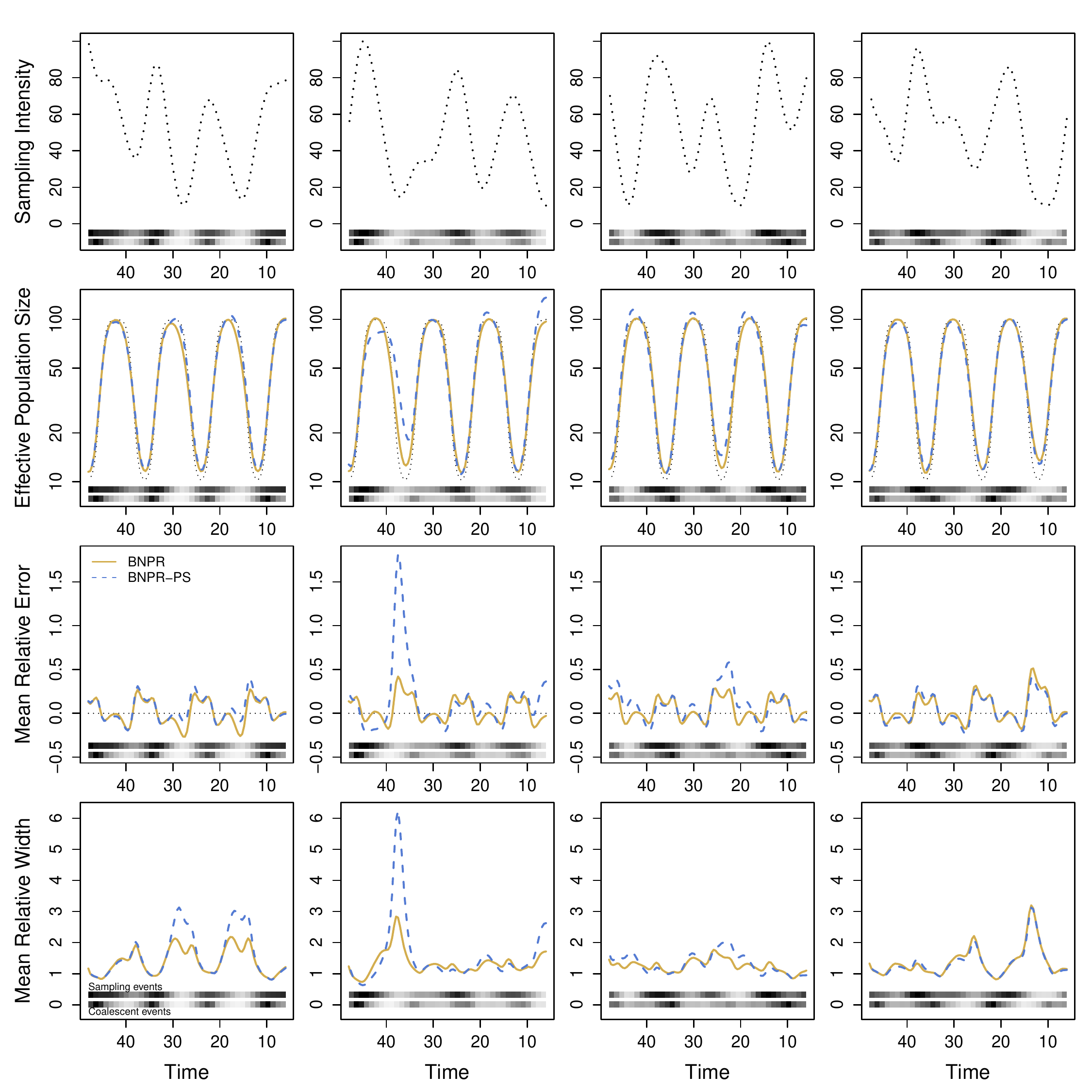}
	\caption{\small Comparison of pointwise statistics for a randomly generated Gaussian process sampling intensity independent of effective population size.
		Visuals as in Fig. \ref{fig:PCComparison}.}
	\label{fig:GPComparison}
\end{figure}

In the negative control simulations section above, we seek to differentiate between misspecification error due to preferential sampling and error due to few observations (long periods without coalescent events). 
In Fig. \ref{fig:PCComparison} we simulated sampling times from random piecewise constant sampling intensities independent from effective population size.
In Fig. \ref{fig:GPComparison} we simulated sampling times from a Gaussian process sampling intensity, also independent from effective population size.
Under BNPR, we see relative errors and relative widths less severe than in the case of preferential sampling as we see in Fig. \ref{fig:Single}, \ref{fig:Comparison},  and \ref{fig:HighBetaComparison}.
However, the performance of BNPR-PS suffers due to the wildly changing ratios of sampling intensity and effective population size.

\subsection*{Regional influenza}

\begin{figure}[htbp]
	\centering
		\includegraphics[width=1.0\textwidth]{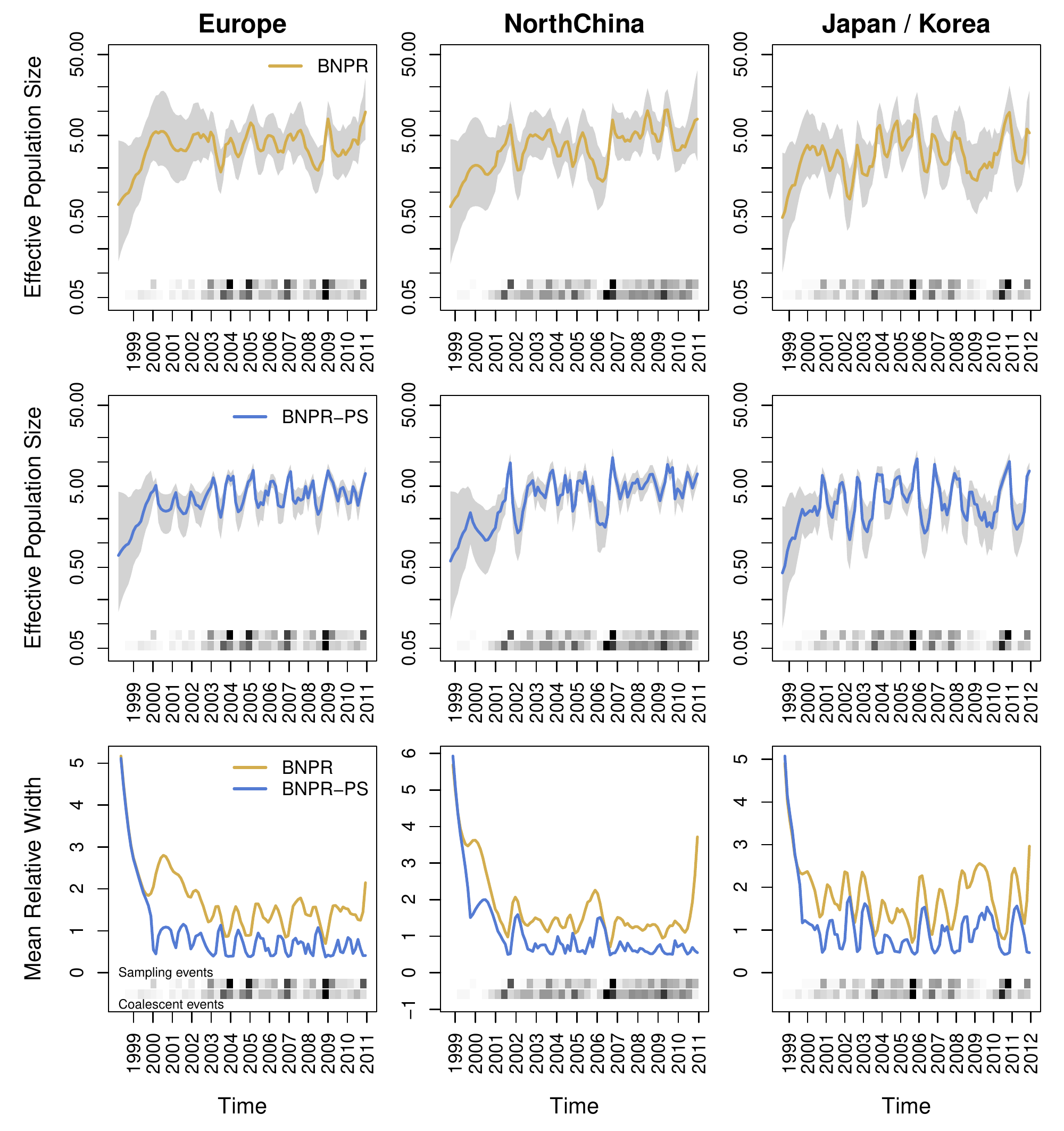}
	\caption{\small BNPR and BNPR-PS models applied to the genealogies inferred from the regional influenza example. We see moderate correlation between effective population size $N^{\boldsymbol{\gamma}}(t)$ and sampling frequencies in the data (Table \ref{tab:RealWorld}). We see improvements in Bayesian credible interval widths, and BNPR-PS performs as well or better than BNPR everywhere in these examples.}
	\label{fig:RegionalInfluenza2}
\end{figure}

\begin{figure}[htbp]
	\centering
		\includegraphics[width=1.0\textwidth]{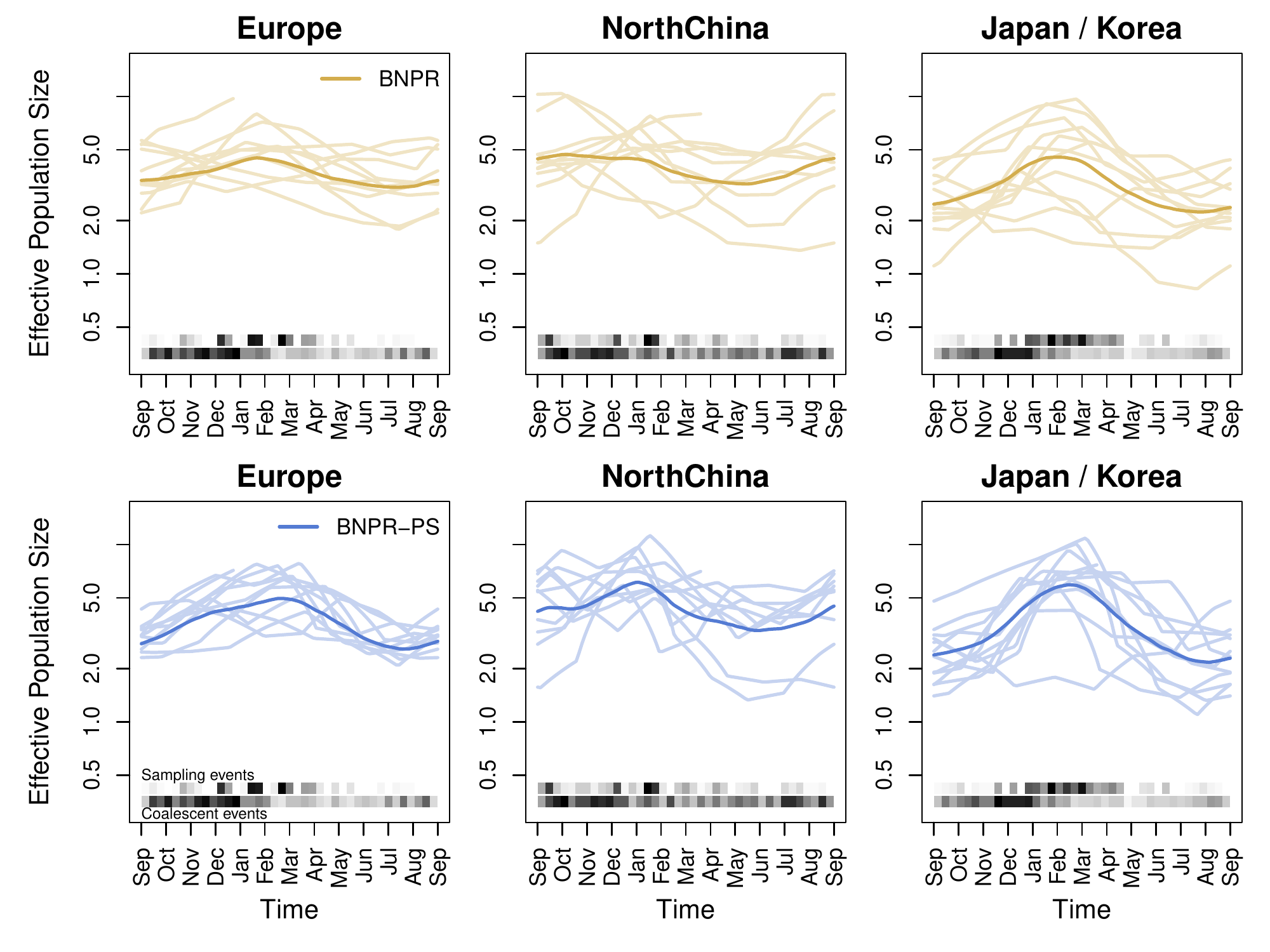}
	\caption{\small BNPR and BNPR-PS models applied to the genealogies inferred from the regional influenza example with years overlaid. We see more pronounced seasonality in the BNPR-PS plots.}
	\label{fig:RegionalSeasonality2}
\end{figure}

\begin{figure}[htbp]
	\centering
		\includegraphics[width=1.0\textwidth]{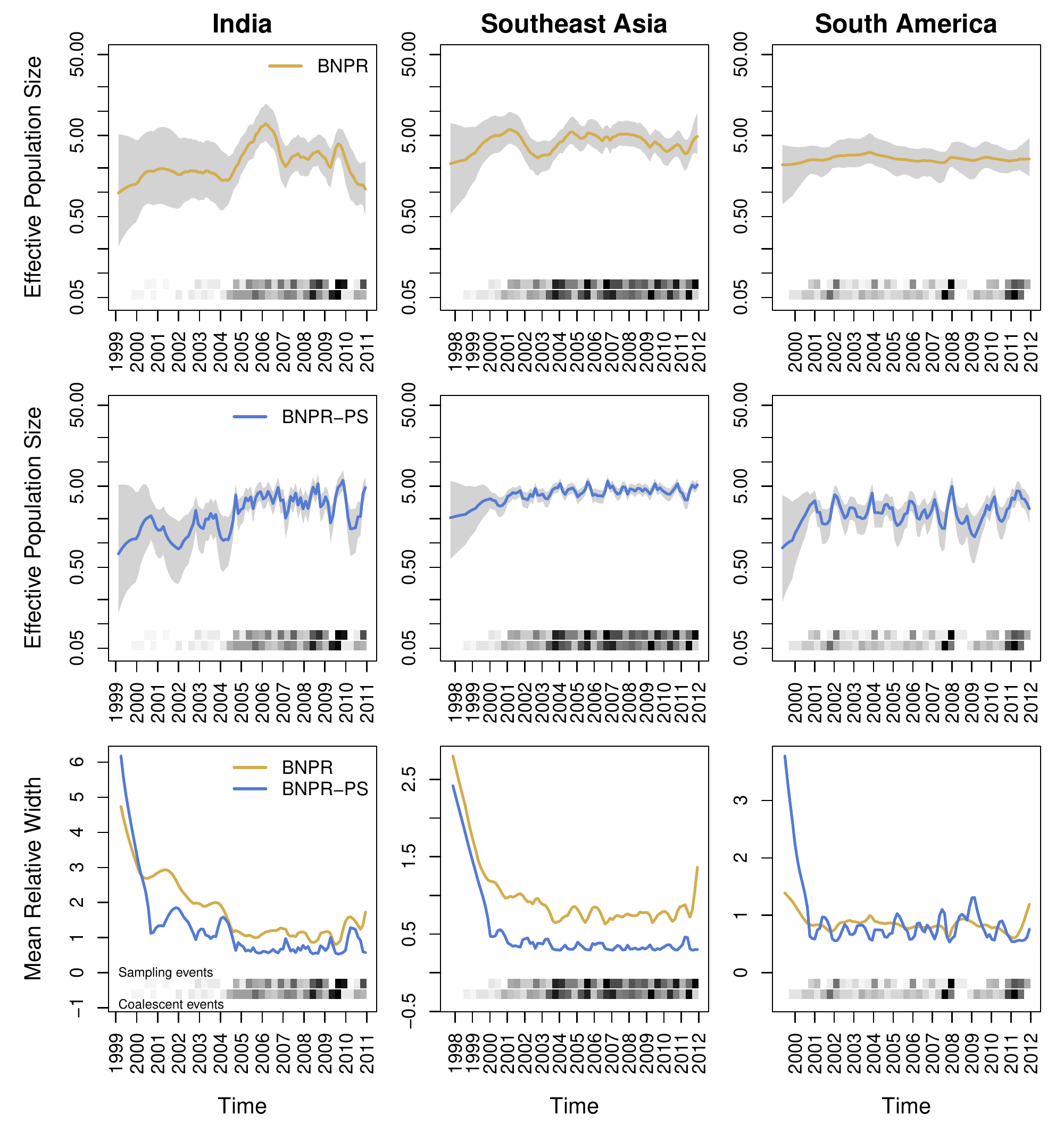}
	\caption{\small BNPR and BNPR-PS models applied to the genealogies inferred from the regional influenza example. Visuals as in Fig. \ref{fig:RegionalInfluenza2}. In South America, we see moderate correlation between effective population size $N^{\boldsymbol{\gamma}}(t)$ and sampling frequencies in the data (Table \ref{tab:RealWorld}). We see improvements in Bayesian credible interval widths, and BNPR-PS performs as well or better than BNPR everywhere in these examples.}
	\label{fig:RegionalInfluenza3}
\end{figure}

\begin{figure}[htbp]
	\centering
		\includegraphics[width=1.0\textwidth]{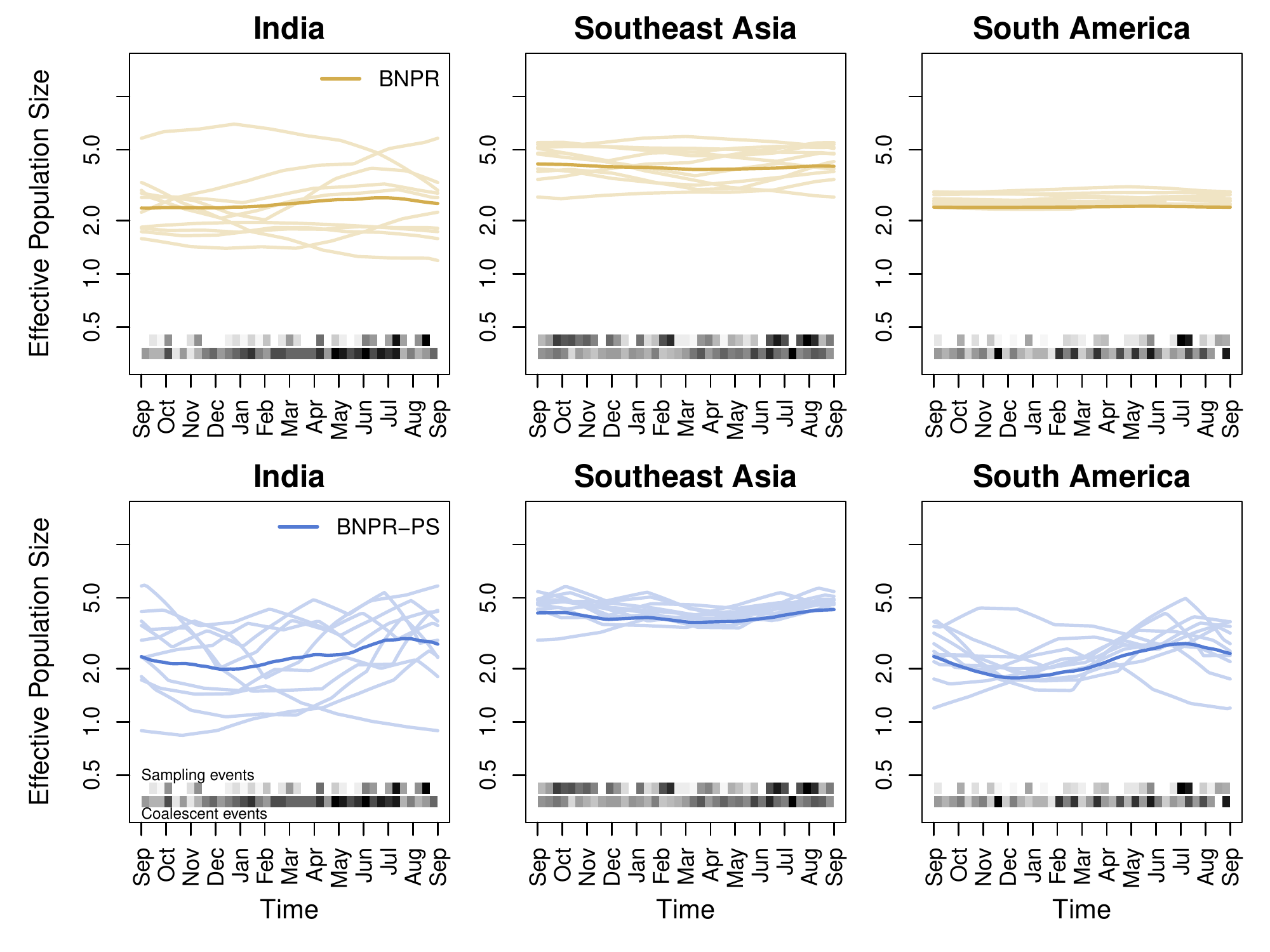}
	\caption{\small BNPR and BNPR-PS models applied to the genealogies inferred from the regional influenza example with years overlaid. We see more pronounced seasonality in the BNPR-PS plots.}
	\label{fig:RegionalSeasonality3}
\end{figure}

We examine three of the remaining regions more closely in Fig. \ref{fig:RegionalInfluenza2} and \ref{fig:RegionalSeasonality2} and the final three regions in Fig. \ref{fig:RegionalInfluenza3} and \ref{fig:RegionalSeasonality3}.
We see much more pronounced seasonality in the estimated effective population size trajectories produced by BNPR-PS,
as well as noticeable improvements in the relative widths of the Bayesian credible intervals.
We see three regions with unusual results.
Whereas most of the regions show some seasonality under BNPR which becomes more visible under BNPR-PS,
India and Southeast Asia both have little seasonality under both methods.
Furthermore, South America has little seasonality under BNPR but some seasonality appears under BNPR-PS. We suspect that these results may be due to inclusion of countries with
different flu seasonal patterns into this region.

\end{document}